\documentclass[sigplan,authorversion,nonacm]{acmart}

\AtBeginDocument{%
  \providecommand\BibTeX{{%
    \normalfont B\kern-0.5em{\scshape i\kern-0.25em b}\kern-0.8em\TeX}}}

\usepackage[binary-units=true]{siunitx} 

\usepackage{xcolor}
\usepackage{ifthen}
\usepackage{tikz}
\usepackage{pgfplots,pgfplotstable}
\usepackage{natbib}
\usepackage[T1]{fontenc}
\usepackage{multirow}

\usetikzlibrary{calc, patterns, spy}

\usepackage{listings}
\AtBeginDocument{\DeclareCaptionSubType{lstlisting}}
\usepackage[frozencache,cachedir=.]{minted}
\usepackage{listings}
\setminted{fontsize=\small,baselinestretch=1}
\usepackage{booktabs}
\usepackage{mdframed}
\usepackage{subcaption}

\definecolor{pairedOneLightBlue}{HTML}{a6cee3}
\definecolor{pairedTwoDarkBlue}{HTML}{1f78b4}
\definecolor{pairedThreeLightGreen}{HTML}{b2df8a}
\definecolor{pairedFourDarkGreen}{HTML}{33a02c}
\definecolor{pairedFiveLightRed}{HTML}{fb9a99}
\definecolor{pairedSixDarkRed}{HTML}{e31a1c}
\definecolor{butter1}{rgb}{0.988,0.914,0.310}
\definecolor{butter2}{rgb}{0.929,0.831,0.000}
\definecolor{butter3}{rgb}{0.769,0.627,0.000}
\definecolor{orange1}{rgb}{0.988,0.686,0.243}
\definecolor{orange2}{rgb}{0.961,0.475,0.000}
\definecolor{orange3}{rgb}{0.808,0.361,0.000}
\definecolor{chocolate1}{rgb}{0.914,0.725,0.431}
\definecolor{chocolate2}{rgb}{0.757,0.490,0.067}
\definecolor{chocolate3}{rgb}{0.561,0.349,0.008}
\definecolor{chameleon1}{rgb}{0.541,0.886,0.204}
\definecolor{chameleon2}{rgb}{0.451,0.824,0.086}
\definecolor{chameleon3}{rgb}{0.306,0.604,0.024}
\definecolor{skyblue1}{rgb}{0.447,0.624,0.812}
\definecolor{skyblue2}{rgb}{0.204,0.396,0.643}
\definecolor{skyblue3}{rgb}{0.125,0.290,0.529}
\definecolor{plum1}{rgb}{0.678,0.498,0.659}
\definecolor{plum2}{rgb}{0.459,0.314,0.482}
\definecolor{plum3}{rgb}{0.361,0.208,0.400}
\definecolor{scarletred1}{rgb}{0.937,0.161,0.161}
\definecolor{scarletred2}{rgb}{0.800,0.000,0.000}
\definecolor{scarletred3}{rgb}{0.643,0.000,0.000}
\definecolor{aluminium1}{rgb}{0.933,0.933,0.925}
\definecolor{aluminium2}{rgb}{0.827,0.843,0.812}
\definecolor{aluminium3}{rgb}{0.729,0.741,0.714}
\definecolor{aluminium4}{rgb}{0.533,0.541,0.522}
\definecolor{aluminium5}{rgb}{0.333,0.341,0.325}
\definecolor{aluminium6}{rgb}{0.180,0.204,0.212}

\definecolor{blind_safe_one_scheme_three_colors}{RGB}{102,194,165}
\definecolor{blind_safe_two_scheme_three_colors}{RGB}{252,141,98}
\definecolor{blind_safe_three_scheme_three_colors}{RGB}{141,160,203}

\definecolor{blind_safe_one_scheme_four_colors}{RGB}{166,206,227}
\definecolor{blind_safe_two_scheme_four_colors}{RGB}{31,120,180}
\definecolor{blind_safe_three_scheme_four_colors}{RGB}{178,223,138}
\definecolor{blind_safe_four_scheme_four_colors}{RGB}{51,160,44}

\definecolor{blind_safe_one_scheme_five_colors}{RGB}{240,249,232}
\definecolor{blind_safe_two_scheme_five_colors}{RGB}{186,228,188}
\definecolor{blind_safe_three_scheme_five_colors}{RGB}{123,204,196}
\definecolor{blind_safe_four_scheme_five_colors}{RGB}{67,162,202}
\definecolor{blind_safe_five_scheme_five_colors}{RGB}{8,104,172} 

\definecolor{blind_safe_one_scheme_seven_colors_grnblu}{RGB}{240,249,232}
\definecolor{blind_safe_two_scheme_seven_colors_grnblu}{RGB}{204,235,197}
\definecolor{blind_safe_three_scheme_seven_colors_grnblu}{RGB}{168,221,181}
\definecolor{blind_safe_four_scheme_seven_colors_grnblu}{RGB}{123,204,196}
\definecolor{blind_safe_five_scheme_seven_colors_grnblu}{RGB}{78,179,211}
\definecolor{blind_safe_six_scheme_seven_colors_grnblu}{RGB}{43,140,190}
\definecolor{blind_safe_seven_scheme_seven_colors_grnblu}{RGB}{8,88,158}

\definecolor{blind_safe_one_scheme_seven_colors}{RGB}{118,42,131}
\definecolor{blind_safe_two_scheme_seven_colors}{RGB}{175,141,195}
\definecolor{blind_safe_three_scheme_seven_colors}{RGB}{231,212,232}
\definecolor{blind_safe_four_scheme_seven_colors}{RGB}{247,247,247}
\definecolor{blind_safe_five_scheme_seven_colors}{RGB}{217,240,211}
\definecolor{blind_safe_six_scheme_seven_colors}{RGB}{127,191,123}
\definecolor{blind_safe_seven_scheme_seven_colors}{RGB}{27,120,55}

\definecolor{yellow_one}{RGB}{255,255,212}
\definecolor{yellow_two}{RGB}{254,217,142}
\definecolor{yellow_three}{RGB}{254,153,41}
\definecolor{yellow_four}{RGB}{217,95,14}
\definecolor{yellow_five}{RGB}{153,52,4}

\definecolor{new_one_seven_colors}{HTML}{ccebc5}
\definecolor{new_two_seven_colors}{HTML}{a8ddb5}
\definecolor{new_three_seven_colors}{HTML}{7bccc4}
\definecolor{new_four_seven_colors}{HTML}{4eb3d3}
\definecolor{new_five_seven_colors}{HTML}{2b8cbe}
\definecolor{new_six_seven_colors}{HTML}{0868ac}
\definecolor{new_seven_seven_colors}{HTML}{084081}
\definecolor{codegray}{rgb}{0.5,0.5,0.5}
\definecolor{lightblue}{HTML}{deebff}

\usepackage{ifthen}
\usepackage{newtxtext,newtxmath}
\usepackage{xcolor}

\newboolean{showcomments}
\setboolean{showcomments}{false}

\makeatletter
\newcommand{\mynote}[3]{%
  \ifthenelse{\boolean{showcomments}}{%
   \fbox{\bfseries\sffamily\scriptsize#1}%
   {\small$\blacktriangleright$\textsf{\emph{\color{#3}{#2}}}$\blacktriangleleft$}}%
  {%
   \@bsphack
   \@esphack
  }%
}
\makeatother

\definecolor{asparagus}{rgb}{0.53, 0.66, 0.42}
\newcommand{\rebuttal}[1]{\textcolor{black}{#1}}

\definecolor{brick}{rgb}{0.8, 0.25, 0.33}

\newcommand{\rev}[1]{\rebuttal{#1}}

\newcommand{\cname}{CINM}
\newcommand{\cinm}{\texttt{cinm}}
\newcommand{\cnm}{\texttt{cnm}}
\newcommand{\cim}{\texttt{cim}}
\newcommand{\memrstr}{\texttt{memristor}}
\newcommand{\upmem}{\texttt{upmem}}

\usepackage{pifont}
\newcommand{\cmark}{\ding{51}}%
\newcommand{\xmark}{\ding{55}}%

\makeatletter
\lstdefinelanguage{mlir}{
    classoffset=0,
    morekeywords={
        module,
        func,
        cinm,
        cnm,
        cim
    },
    morestring=[b]",
    alsoletter={\%},
    keywordsprefix={\%}
}
\makeatother

\begin{document}
%
\title[CINM]{CINM (Cinnamon): A \underline{C}ompilation Infrastructure for Heterogeneous Compute \underline{I}n-Memory and Compute \underline{N}ear-\underline{M}emory Paradigms}
\date{}



 \author{Asif Ali Khan}
 \affiliation{%
   \institution{TU Dresden}
   \city{Dresden}
   \country{Germany}}
 \email{asif_ali.khan@tu-dresden.de}
 \orcid{0000-0002-5130-9855}

 \author{Hamid Farzaneh}
 \affiliation{%
   \institution{TU Dresden}
   \city{Dresden}
   \country{Germany}}
 \email{hamid.farzaneh@tu-dresden.de}
 \orcid{0000-0002-1780-6217}

 \author{Karl F. A. Friebel}
 \affiliation{%
   \institution{TU Dresden}
   \city{Dresden}
   \country{Germany}}
 \email{karl.friebel@tu-dresden.de}
 \orcid{0000-0001-9534-3978}

 \author{Clément Fournier}
 \affiliation{%
   \institution{TU Dresden}
   \city{Dresden}
   \country{Germany}}
 \email{clement.fournier@tu-dresden.de}
 \orcid{0000-0002-5661-3004}

 \author{Lorenzo Chelini}
 \affiliation{%
   \institution{Intel Switzerland}
   \city{Zurich}
   \country{Switzerland}}
 \email{lorenzo.chelini@intel.com}
\orcid{0000-0001-8539-2397}
\author{Jeronimo Castrillon}
\affiliation{%
  \institution{TU Dresden, ScaDS.AI, and \\ Barkhausen Institut}
   \city{Dresden}
   \country{Germany}}
 \email{jeronimo.castrillon@tu-dresden.de}
 \orcid{0000-0002-5007-445X}


%
%
\renewcommand{\shortauthors}{Khan et al.}

\begin{abstract}
  The rise of data-intensive applications exposed the limitations of conventional
  processor-centric von-Neumann architectures that struggle to meet the off-chip
  memory bandwidth demand.  Therefore, recent innovations in computer
  architecture advocate compute-in-memory (CIM) and compute-near-memory (CNM),
  non-von-Neumann paradigms achieving orders-of-magnitude improvements in performance
  and energy consumption.  Despite significant technological breakthroughs in the last few
  years, the programmability of these systems is still a serious challenge.
  Their programming models are too low-level
  and specific to particular system implementations.  
  Since such future architectures are predicted to be
  highly heterogeneous, developing novel compiler abstractions and frameworks
  becomes necessary.  To this end, we present \emph{CINM (Cinnamon)}, a first 
  end-to-end compilation flow that
  leverages the hierarchical abstractions to generalize over different CIM and CNM
  devices and enable device-agnostic and device-aware optimizations.  Cinnamon
  progressively lowers input programs and performs optimizations at each level in
  the lowering pipeline.  To show its efficacy, we evaluate CINM
  on a set of
  benchmarks for a real CNM system (UPMEM) and the memristors-based CIM
  accelerators. We show that Cinnamon, supporting multiple hardware
  targets, generates high-performance code comparable to or better than
  state-of-the-art implementations.
\end{abstract}


\maketitle



\sloppy
\section{Introduction}
\label{sec:intro}
Application domains such as social and streaming media, internet-of-everything,
communications and services, and virtual assistant technologies such as Alexa
and Siri are generating data at a break-neck pace, i.e., in the
\emph{quintillion bytes} range every day.  This mind-boggling data volume  is
mostly raw and requires processing and analysis~\cite{datagrowth}. 
In the conventional \emph{processor centric} von-Neumann computing paradigm,
these applications quickly hit hard performance and energy efficiency
boundaries as data have to be moved between the CPU and the memory via a narrow
memory channel.  On a mobile device, the data movement alone consumes 62\% of
the total system energy~\cite{data_movement_energy}. 
To overcome this data movement and other challenges associated with the memory 
subsystem, computer
architects are moving to \emph{non-Von-Neumann} system models like
\emph{computing near memory} (CNM)~\cite{cnm-review} and \emph{computing in
memory} (CIM)~\cite{cim-review}.\footnote{CIM and CNM are also referred to as (PIM, IMC) and (NMP, PNM), respectively, in the literature. We will use CIM and CNM in this paper.}  The idea is to bring computations
closer to the data. 
In CNM, dedicated CMOS logic is integrated into the memory chip to diminish the
data movement problem.  Conceptually, this tight coupling of the logic and
memory devices can be applied at any level in the memory hierarchy with various
memory technologies.  For DRAM, both planar and stacked structures, such as
Micron's hybrid memory cube~\cite{hmc}, AMD's and SK Hynix's high bandwidth
memory~\cite{hbm}, and Samsung's wide I\/O~\cite{wideio} have been used to
realize CNM systems~\cite{cnm-review}.  While CNM greatly reduces the data
movement on the CPU bus, 
it still requires communicating data between
the memory and the compute units.  The CIM model completely eliminates data
movement to compute units by exploiting the physical properties of the memory devices to
implement various logic and compute operations in-place~\cite{cim-review}.  
CIM systems based on novel memory devices with inherent computing capabilities, such
as phase change memory (PCM), resistive RAM (RRAM), magnetic RAM (MRAM), and
spintronics-based racetrack memories (RTMs) have demonstrated orders of
magnitude performance and energy gains for machine learning and other
application domains~\cite{memristor, mehonic2020memristors,
ielmini2018memory, cim-mram, blasing2020magnetic}.    

Of late, several innovative CNM and CIM systems have been proposed, and some of them are \rev{even 
commercially available~\cite{cinm-review}}. 
These include domain-specific architectures such as the Neurocube~\cite{neurocube}, 
ISAAC~\cite{isaac}, Microsoft Brainwave NPU~\cite{BrainwaveNPU}, 
and several DNN accelerators~\cite{MLAccelerators} among others. 
These systems are orders of magnitude faster 
and more energy-efficient than general-purpose Von-Neumann machines,
but only target specific application domains. 
UPMEM~\cite{UPMEM} has shown case studies of CNM in more general-purpose off-the-shelf systems. 
Recently, Samsung~\cite{samsungPIM, samsungPIMISSCC} and 
SK hynix~\cite{skhynix} proposed machine learning specific CNM 
systems based on the HBM2 and GDDR6 DRAM standards supporting TFLOPS. 
On the CIM front, in just the last couple of years, all major memory 
tech giants such as Samsung~\cite{jung2022crossbar}, 
TSMC~\cite{chiu202222nm, khwa202240nm}, Intel~\cite{intel-sram-cim}, GlobalFoundries~\cite{gf-fefet-crossbar}, and 
IBM~\cite{khaddam2022hermes, ibm-64-cores} have 
fabricated CIM chips based on memristive and CMOS 
technologies that attain unparalleled performance 
and energy efficiency. 

Even though various companies currently provide CIM and CNM systems for machine learning and other application domains (such as Axelera, d-Matrix, Synthara, UPMEM, etc.), their programmability remains a significant challenge.
Most of these systems provide low-level device libraries and leave the
mapping problem, synchronization, and optimizations to the programmer. This makes the
programmability and operability of these devices extremely difficult. In isolated efforts, 
compilers have been proposed to automatically map compute
primitives to devices and perform load balancing and technology-specific
optimizations~\cite{occ, cnm-compiler, tom-cnm}. 

However, they
target only homogeneous architectures and are application-specific, 
e.g., GEMM on memristive crossbars~\cite{occ}.
Since future architectures are predicted to be highly heterogeneous and general-purpose, there is a pressing
need to develop novel compiler abstractions and compiler frameworks that enable device-agnostic and device-specific optimizations. The same is also highlighted by several recent articles including one from Meta (Facebook) \rev{which} states: ``We’ve investigated applying processing-in-memory (PIM) to our workloads and determined there are several challenges to using these approaches. Perhaps the biggest challenge of PIM is its programmability''~\cite{meta-ai-accel}. 

To this end, our goal is to develop a high-level framework that abstracts over CIM and CNM devices, enabling their programming through high-level frameworks and domain-specific languages, and generating highly efficient code for them. 
We present CINM, pronounced as \emph{Cinnamon}, a novel
framework based on the \emph{multi-level intermediate representation}
(MLIR) that empowers the progressive lowering of abstractions and allows reasoning about computational primitives and their memory behavior and operations at various abstractions. CINM supports PCM and RRAM-based CIM accelerators and the UPMEM CNM architecture.\footnote{The selection of architectures is influenced by the availability of infrastructure where these systems can be
evaluated.}

We are using a high-end real UPMEM system and the extended gem5 simulator~\cite{occ} to evaluate our generated codes for CNM and CIM systems, respectively. For PCM and ReRAM-based accelerators, \cname{} implements and extends OCC~\cite{occ}, an automatic compilation flow for memristive crossbar arrays. 
The hierarchical lowering in the CINM enables identifying the
most suitable target for each primitive in the input application and 
transformations at different abstractions to optimize for individual devices. 
For evaluation, we use the sets of benchmarks available for these systems, i.e., PrIM benchmarks for CNM~\cite{cnm-benches} and the machine learning benchmarks from~\cite{occ} for CIM. Concretely, we make the following contributions: 

\begin{enumerate}
    \item We investigate the landscape of CIM and CNM systems to understand their properties and compute \rev{the} primitives they support (Section~\ref{subsec:motiv}).
    \item We present CINM, an end-to-end compilation framework based on MLIR that seamlessly maps computational motifs to different backend targets (Section~\ref{subsec:overview}).
    \item \cname{} implements multiple hardware-oblivious and hardware-specific abstractions. Concretely, we introduce \emph{cim}/\emph{cnm} abstractions that implement abstract operations for CIM/CNM paradigms, which are subsequently lowered differently for different hardware targets in their respective device dialects (Section~\ref{subsec:lowering}). 
    \item We introduce a high-level \cinm{} dialect that abstracts over all CINM devices and provides a placeholder for implementing cost models to automate \rev{the} mapping of $k$ kernels/regions onto $d$ devices in a heterogeneous system setup (Section~\ref{sec:cinm}).
    \item We implement device-specific abstractions to perform device-aware optimizations and mapping to their respective libraries. 
    \item Our evaluation shows that \cname{} can effectively reproduce or beat the performance of the hand-optimized codes in the selected benchmark suites (Section~\ref{sec:eval}).
\end{enumerate}


\section{Background and motivation}
\label{sec:background}
This section presents MLIR as well as the CNM and CIM computing models using
different memory technologies. 

\subsection{The MLIR compiler infrastructure}
\label{subsec:MLIR}
MLIR is a toolkit to represent and transform intermediate representation (IR)
at different abstraction levels across different application domains and
heterogeneous hardware targets~\cite{mlir}.  It offers a nonopinionated IR with
few builtins, leaving most of the IR customizable. MLIR allows compiler
developers to plug in into the compiler their own abstraction and empowers them
to optimize for a specific domain or target by matching at the appropriate
abstraction levels. 

MLIR implements a set of reusable abstractions modeled with \emph{dialects}.  A
dialect is a logical group of custom types, operations, and attributes.
Operations are building blocks of the IR and consume and produce new values.
Each value in MLIR is associated with a type known at compile time.  Attributes
associate compile-time information to operations.  Dialects in MLIR preserve
transformation validity preconditions in their IR in order to minimize the cost
and complexity of analysis passes.  They are typically associated with domains
(\texttt{linalg} with linear algebra, \texttt{TOSA} with tensor operations),
representations (\texttt{affine} with the polyhedral model, \texttt{scf} with
control flow), or targets (\texttt{gpu}, \texttt{cim}).  Abstractions in MLIR
can be progressively lowered (e.g., from high-level domain-specific to
low-level platform-specific dialects) and raised~\cite{prograising}.

\subsection{Compute near memory}
\label{subsec:cnm}
Compute near memory (CNM) is a data-centric paradigm aiming to process data
in memory proximity.  Compute units, e.g., CPU, GPU, FPGA, ASIC, or CGRA, are
physically placed closer to the memory (in the memory controller, in
peripheral circuitry, on the memory chip, or connected to the memory chip via a
shared crossbar) to minimize data movement. \rev{The original idea of CNMs dates back to 70s~\cite{stone1970logic}}. In the 90s, architectures such as EXECUBE~\cite{kogge1994execube} and
IRAM~\cite{patterson1997iram} demonstrated significant performance gains in a
range of applications.  However, design complexity and manufacturing costs
hindered commercialization.  Recent advances in manufacturing and stacking
technologies alleviate these practicality concerns, paving the way for many
novel CNM architectures. 

Stacked DRAM structures such as the hybrid memory cube (HMC)~\cite{hmc} and the 
high bandwidth memory (HBM)~\cite{hbm} are considered the true enablers of CNM systems. 
These architectures stack multiple DRAM dies on top of a logic layer using 
through silicon vias (TSVs), where the logic layer can implement fixed function units. 
These stacked solutions deliver higher bandwidth and improved performance 
compared to other DRAM families but can lead to higher refresh power and limited capacities. 
UPMEM integrated co-processors with the DDR4 DRAM on the same DRAM die~\cite{UPMEM}. 
The co-processor, known as the data processing unit (DPU), 
is a general-purpose 32-bit RISC processor. 
Due to its massive local and cumulative bandwidth and parallelism, 
UPMEM demonstrated an order of magnitude gains in performance and energy 
consumption on different applications~\cite{cnm-benches}. 

Each DPU has a small private scratchpad working RAM (WRAM) backed by the shared main RAM (MRAM). 
UPMEM provides an SDK and a set of tools that allow developers to adapt to the PIM programming. 
More recently, Samsung and SK Hynix presented their FIMDRAM~\cite{samsungPIM, samsungPIMISSCC} 
and AiM~\cite{skhynix} architectures, respectively. 
Similar to UPMEM, these architectures integrate co-processors on the 
same DRAM die (using HBM2 and GDDR6 DRAMs). 
However, unlike UPMEM, the co-processors in both architectures are 
optimized explicitly for ML-specific workloads. 


\subsection{Compute in memory}
\label{subsec:cim}
The compute in memory paradigm radically departs from traditional architectures by 
implementing certain compute motifs \rev{in memory} using the physical attributes of the devices. 
Memristive devices such as PCM and RRAM cells can be programmed to different resistance 
states using external current/voltage, where each state represents some information. 
When organized in a crossbar configuration, these 
memristive devices allow for in-place fixed-size matrix-vector (MV) multiplication in 
constant time~\cite{memristorApplication}. 
However, these computations are in the analog domain and require converters from the 
digital to the analog domain and back. 
In a different crossbar setup, memristors can be used to implement the entire set of 
logical operations~\cite{kvatinsky2013memristor} entirely in the 
digital domain. 
The write operation in these resistive technologies is typically very slow and reduces 
the device's lifetime. Therefore, the selection of an application for CIM 
acceleration requires careful consideration.  

Magnetic memories such as MRAM and RTM can also be used to implement certain operations in place. 
The tunnel magnetoresistance in the magnetic tunnel junctions (MTJs) of MRAM cells is a natural implementation of the XOR operation, which can be exploited to implement other logic operations~\cite{cim-mram}. 
Similar to memristors, MRAM cells can also be organized in crossbars to realize MV operations~\cite{jung2022crossbar}. 
RTM devices also use MTJs as access interfaces and can use the same basic principles to implement various logic operations~\cite{RTMAdder}. 
They also offer novel access mechanisms that allow efficient implementation of population count and the majority operations~\cite{ollivier2021pirm, hdc}. Conventional charge-based SRAM and DRAM technologies can also implement a series of logic and compute operations in-place~\cite{ambit, XSRAM}.


\subsection{The need for CIM and CNM abstractions}
\label{subsec:motiv}
\begin{figure}[tbh]
\includegraphics[width=0.96\columnwidth]{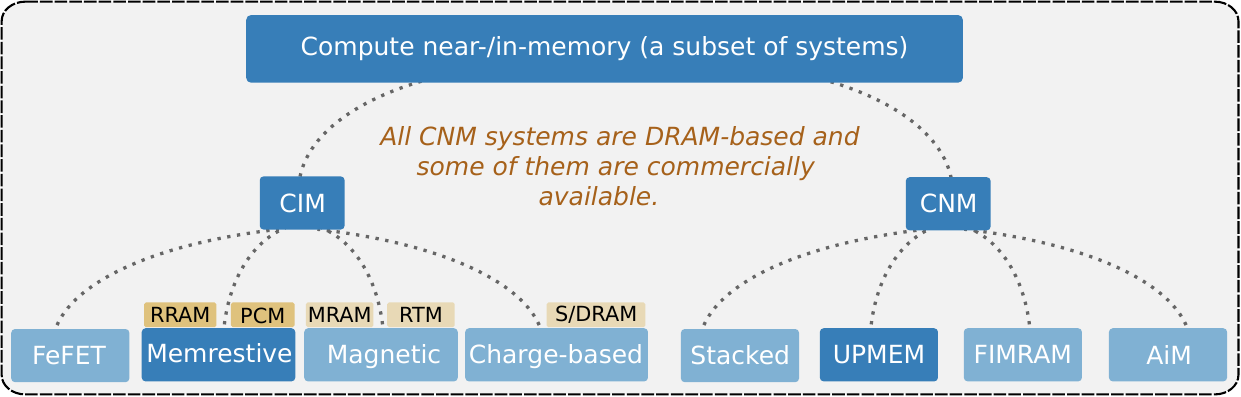}
\caption{A partial taxonomy of CNM and CIM systems based on pronounced technologies.}
\label{fig:taxonomy}
\end{figure}

\autoref{fig:taxonomy} presents a partial taxonomy of prominent CNM and CIM systems. 
On the CNM side, the figure shows only real-world systems. 
Similarly, only promising and mature CIM technologies are presented on the CIM side. 

\begin{figure}[tbh]
\includegraphics[width=0.85\columnwidth]{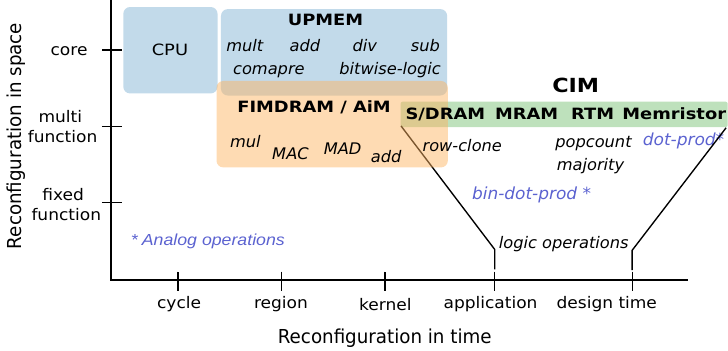}
\caption{CNM and CIM programmability landscape.
}
\label{fig:cinm_ops}
\end{figure}

These architectures are typically optimized for a specific domain or function. \autoref{fig:cinm_ops} shows the landscape of these architectures with their supported operators and their specificity and flexibility to be reconfigured in time and space. For instance, the general-purpose CPU is programmed at the granularity of the core every new instruction cycle. On the contrary, application-specific ICs (ASICs) are optimized for a particular application and can not be reprogrammed. The near-memory logic in CNM systems can be general-purpose (UPMEM), or multi-function (AiM, FIMDRAM), and they are programmed at the kernel and region granularities where, in the latter case, a kernel is partitioned into regions before offloading it to the CNM devices. CIM systems are usually fixed-function (e.g., in dot-product), but they can also be multi-function (e.g., logic operations) and can be programmed at the application granularity.

Unfortunately, even for this limited set of systems, there is a lack of programming models 
that abstract over them and can be leveraged to program them. 
All CNM and CIM systems use low-level, architecture-specific libraries to expose their device traits. 
The radically different design decisions and architectures of these systems make their 
programmability a serious challenge. For instance, in UPMEM, 
the programmer is responsible for load balancing on thousands of DPUs, 
explicit data movement and bandwidth management 
between the CPUs and DPUs, MRAM and WRAM, and the data coherency~\cite{UPMEM, cnm-benches}. 
The AiM architecture has unique features that allow operations including row clone, 
element-wise multiplication, 
and addition on a set of banks with different granularities~\cite{skhynix}. 
However, it is not clear how these systems are programmed. 
Samsung's FIMDRAM has its own \emph{closed-source} software stack~\cite{samsungPIM}. 
The programming models and tools of different CIM technologies, e.g., for PCM~\cite{occ}, 
do not apply to other technologies (such as MRAM) as they have different properties.

\begin{figure}[tbh]
\begin{subfigure}{\columnwidth}
\centering
\begin{minted}[breaklines=true, framesep=2mm, frame=lines, fontsize=\tiny\ttfamily, ]{c}
BARRIER_INIT(my_barrier, NR_TASKLETS);
int main() { ...
    barrier_wait(&my_barrier);
    int32_t point_per_tasklet = (ROWS*COLS)/NR_TASKLETS;
    uint32_t mram_base_addr_A = (uint32_t) (DPU_MRAM_HEAP_POINTER ); 
    uint32_t mram_base_addr_B = (uint32_t) (DPU_MRAM_HEAP_POINTER + 
        ROWS * COLS * sizeof(T)); 
    uint32_t mram_base_addr_C = (uint32_t) (DPU_MRAM_HEAP_POINTER + 
        2 * ROWS * COLS * sizeof(T)); 
    for(int i = (tasklet_id * point_per_tasklet) ; 
        i < ( (tasklet_id+1)*point_per_tasklet ) ; i++) {
        if( new_row != row ){ ...  
            mram_read((__mram_ptr void const*) (mram_base_addr_A +
                mram_offset_A), cache_A, COLS * sizeof(T));
        }
        mram_read((__mram_ptr void const*) (mram_base_addr_B + 
            mram_offset_B), cache_B, COLS * sizeof(T));
        dot_product(cache_C, cache_A, cache_B, number_of_dot_products);
    ... } ...
    mram_write( cache_C, (__mram_ptr void *) (mram_base_addr_C 
        + mram_offset_C), point_per_tasklet * sizeof(T));
}
\end{minted}
\caption{Matmul on UPMEM. Each tasklet runs this code to generate a single element in the resultant matrix.}
\label{fig:matmul-upmem}
\end{subfigure}
\hfill
\begin{subfigure}[t]{\columnwidth}
\centering
\begin{lstlisting}[language=mlir,basicstyle=\tiny\ttfamily,linerange={1-4},breaklines=true,postbreak=\mbox{\textcolor{red}{$\hookrightarrow$}\space},frame=tb,linewidth=\columnwidth]
func.func @matmul(%A: tensor<64x64xi32>, %B: tensor<64x64xi32>, %C: tensor<64x64xi32>) -> tensor<64x64xi32> {
    %D = linalg.matmul ins(%A, %B : tensor<64x64xi32>, tensor<64x64xi32>) outs(%C: tensor<64x64xi32>)
    return %D : tensor<64x64xi32>
}
\end{lstlisting}
\caption{GEMM code at the \texttt{linalg} abstraction in \cname{}.}
\label{fig:gemm-linalg}
\end{subfigure}
\caption{Matmul code using the UPMEM programming model and the device-unaware \texttt{linalg} abstraction in \cname{}.}
\label{fig:abstractions}
\end{figure}


\rev{Figure~\ref{fig:matmul-upmem} shows an implementation of the GEMM kernel in UPMEM's C interface. 
Performing simple read and write for a computation requires specific API calls and manual address translation.}
For other architectures, this code has to be completely rewritten using their device-specific function calls.
With these programming models, it is extremely difficult to program and effectively 
utilize heterogeneous systems integrating these technologies. Even for the same system, 
any device or system changes may lead to a considerable rewriting of the input applications. 
To enable the integration and exploration of these devices in heterogeneous setups, 
novel programming models that abstract from these devices to a higher level \rev{are needed}. 
\cname{}'s device-agnostic abstraction is a step in that direction. 
The GEMM input to \cname{} is device independent, as shown in \autoref{fig:gemm-linalg}, 
and can be lowered to CIM or CNM device code. 
This also highlights the expressiveness and conciseness of CINM compared to the low-level device-specific programming models.
\cname{} allows rigorous analysis and 
reasoning about individual kernels and supports a rich 
set of optimizations and device interfaces.

Compared to device libraries, compiler \rev{frameworks} like \cname{}
are interesting alternatives or complements because for most CIM and CNM systems, 
such libraries do not exist or are device-specific and thus not portable.
In addition, libraries use kernels as-is, while compilers like ours, if the device supports it, 
can fuse operations to reduce the data movement. Compiler-optimized codes have also been shown to be on-par 
with the libraries~\cite{gemm-gpu}. 
\cname{} can be extended to map to optimized libraries when they 
become available (in the same way many DSLs map to, e.g., BLAS calls).

         \begin{figure*}[tbh]
        \begin{center}
        \includegraphics[width=\textwidth]{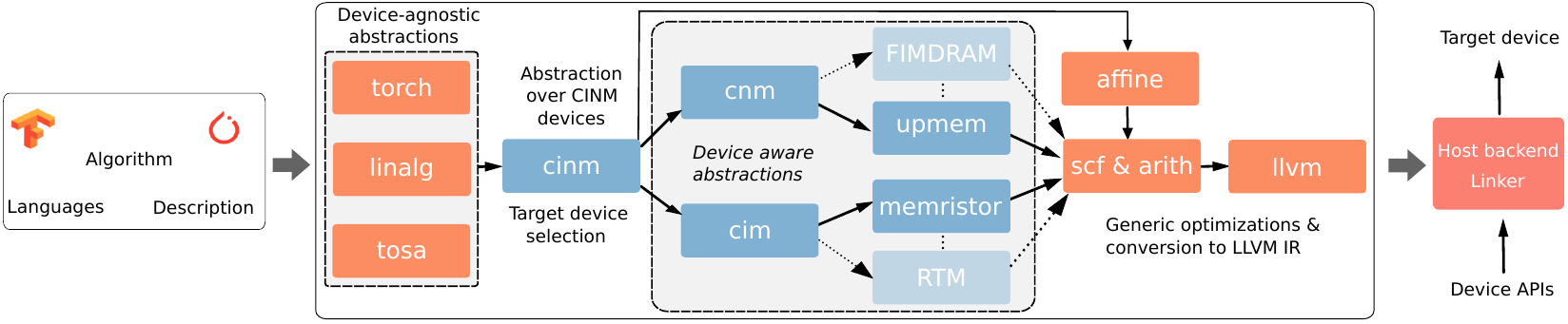}
        \caption{The CINM compiler. The abstraction lowers from left to right. Blue boxes show the new dialects introduced in \cname{}. }
        \label{fig:flow}
        \end{center}
        \end{figure*}


\section{CINM (Cinnamon): Compilation for in and near memory computing}
\label{sec:cinm}
        The explosion of CIM and CNM technologies and architectures has led to
        fragmented toolchains and device-specific low-level programming APIs.  This
        imposes a high barrier of entry for programmers, leading to low adoption and
        potentially slower technology evolution cycles.
        Based on the taxonomy and operator pools described in Section~\ref{subsec:motiv},
        we present
        \cname{} -- an end-to-end compilation flow that generates high-performance code
        for various target devices.  \cname{} leverages MLIR to optimize input programs
        by progressively lowering from high-level domain abstractions to low-level
        device abstractions. Each device dialect supports device-aware transformations
        to ensure effective utilization of the underlying system.  In the scope of this
        work, we target only memristors-based CIM systems and the UPMEM CNM systems, as
        highlighted in \autoref{fig:taxonomy}.  However, considering the ongoing device
        innovations, we design our abstractions with a special focus on extensibility.
        The restriction to selected architectures in this work is enforced by the lack
        of open-source tools for other architectures that can be used to evaluate the
        generated code for them.

        \subsection{The \cname{} lowering pipeline}
        \label{subsec:overview}
        \autoref{fig:flow} presents a high-level overview of the
        \cname{} compilation flow. The entry point to the compilation flow is the
        \texttt{cinm} dialect, which currently accepts the IRs from the \texttt{linalg},
        \texttt{TOSA} and \texttt{torch} abstractions. 
        The \cinm{} abstraction is a
        generalization over different CINM technologies and takes over the shared
        responsibilities of host-device interfacing and device mapping. The latter
        may also require the input program to be (re)written in CINM-supported operations (see
        \autoref{fig:cinm_ops}), which can then be processed by the low-level dialects.
        The \cinm{} dialect is then lowered to the \cim{}, \cnm{} or \texttt{affine} dialects.

        \cim{} and \cnm{} abstractions implement custom types and operations that are
        common to these architectures.  For instance, in all CNM devices, the host
        allocates the grid of compute devices and transfers data before launching the
        kernel.  The \cnm{} dialect implements abstract prototypes of these functions
        using custom types that are contextually converted to the device types.  The concrete
        mapping from \cnm{} operations to the target devices is then accomplished using
        device dialects.  
        These dialects serve as interfaces to their respective accelerators and
        runtimes. All device dialects provide their lowering for code generation,
        which could mean emitting runtime library calls (e.g., for \upmem{}) or CPU
        instructions (for devices embedded as ISA extensions).


        \subsection{Progressive lowering}
        \label{subsec:lowering}
        This section describes the \cname{} pipeline, particularly focusing on our newly added dialects
        and their primitives.

        \subsubsection{\rev{CINM front-ends}}
        \label{subsec:front-end}
        \rev{
        The entry point to the \cname{} framework is the \cinm{} abstraction. To arrive at this abstraction, \cname{} presently starts from the high-level abstractions \texttt{linalg}, \texttt{tosa}, or \texttt{torch}, which serve as widely adopted targets for several DSLs and frameworks, particularly in the ML domain, including TensorFlow~\cite{tensorflow}, ONNX~\cite{onnx}, OpenEarth Compiler~\cite{oec}, CFDlang~\cite{cfdlang-mlir}, and others.
        In terms of applicability, they handle a wider set of applications from the ML and other domains, as demonstrated in Section~\ref{sec:eval}, but can not necessarily handle \textit{any} input application. It is important to note that the \cname framework itself is not tied to these front-ends. In general, the program representation can be in any high-level or domain-specific language that already provides a front-end to the \texttt{linalg} abstraction, e.g., Tensor Comprehension (via teckyl), TorchScript (via torch-mlir), and a subset of C/C++ (via Polygeist~\cite{polygeist}), among others. For non-\texttt{linalg} operations or fine-grained control over the target selection process, front-ends must provide direct lowering into \texttt{cinm}. More advanced front-ends, such as Mojo~\cite{mojo}, which handle optimizations in a compiler-as-library approach, would need to entirely sidestep \texttt{cinm} and directly emit \texttt{cnm} and \texttt{cim}.
        }

        \subsubsection{The \texttt{cinm} dialect}
        \label{sss:cinm}
        The \texttt{cinm} dialect is the entry point to the \cname{} flow and is responsible for target selection, i.e.,
        delegating the implementation of $k$ kernels or regions in an input application to the most suitable $d$ devices. 


        Target selection is not arbitrary, as it requires precise cost models for individual devices and an exhaustive search mechanism to evaluate tradeoffs before making mapping decisions.
        For some operations such as \texttt{cinm.gemm}, estimating the execution time (or other metric) on a specific architecture might not be that difficult.
        \rev{
        However, for more complex operations or a combination of the easy-to-estimate operations, additional analysis and
        rewriting passes becomes necessary.} For instance,
        none of the discussed CINM architectures are optimized for convolution and
        contraction operations. For tensor contractions or convolutions, in the
        absence of rewriting, the subsequent lowering from \texttt{cinm} will have no choice but to map them to more general compute-capable devices, like, e.g.,
        UPMEM or the host CPU. However, having detected these kernels and identified
        them as profitable, they can be rewritten as matrix-matrix multiplications,
        amenable to both CIM and CNM acceleration. Therefore, the \texttt{cinm} abstraction must evaluate all code variants before making any mapping decision.

        \rev{CINM provides a mechanism to implement and integrate device cost models that can be used for target selection and offloading decisions. Due to the absence of cost models for  our targeted devices (see Section~\ref{subsec:costmodel}), CINM presently selects offloading targets as follows. For a given application, the user can optionally specify the target device on the command line. In the absence of user specification, the analysis passes within the \cinm{} abstraction automatically identifies a target device. For instance, matmul-like operations or those that can be rewritten as matmul are greedily offloaded to the CIM crossbar if the tensor dimensions are greater than a user-defined threshold. Convolutions and contractions are identified using the analysis algorithm from OCC~\cite{occ}. Similarly, search operations suited to content-addressable memories (CAMs) can be detected using the analysis algorithm from C4CAM~\cite{c4cam}. For all other \cinm{} operations, CINM presently selects UPMEM as a target. 
        ~\\
        As we explain in Section~\ref{subsec:costmodel}, CINM is designed to support heterogeneous systems and enable offloading at finer granularities. For instance, within a kernel, if some operations can be accelerated on one device and others on another, the framework should seamlessly perform such offloading. However, this requires extensive analysis, which is only possible when the cost models become available. 
        }

        For the cost model to work on this abstraction, it must be aware of all the
        primitives supported by the underlying devices. \texttt{cinm}, therefore,
        exposes the set of operations listed in \autoref{tab:cinm-ops} to be targeted
        during offloading. These are the MLIR counterparts of those found in
        \autoref{fig:cinm_ops}. 
        \rev{\texttt{cinm} IR is emitted by a conversion pass that applies to the \texttt{linalg} input. The \texttt{linalg} input may have been produced by a front-end (see Sec.~\ref{subsec:front-end}) for a high-level framework or another higher-level front-end dialect like \texttt{tosa} or \texttt{torch}.
        \texttt{linalg} operators that do not have a direct counterpart in the \texttt{cinm} dialect are first canonicalized into a sequence of lower-level operators that are mappable to \texttt{cinm}.
        For instance, in an MLP application represented at the \texttt{tosa} abstraction, the \texttt{tosa.fully\_connected} operation is first decomposed into a sequence of \texttt{linalg} operations (\texttt{transpose}, \texttt{matmul}, and bias addition using a \texttt{generic} operation). During \texttt{linalg} to \texttt{cinm} conversion, the \texttt{generic} operation responsible for adding the bias is rewritten with a \texttt{cinm.add} operation in the canonicalization pass.
        Operators that still cannot be converted are run on the host CPU, as they cannot participate in \texttt{cinm} target selection.
        }
        \rev{Figure~\ref{fig:proglowering} shows a convolution kernel at the \texttt{linalg} (Figure~\ref{fig:linalgIR}) and \texttt{cinm} (Figure~\ref{fig:cinmIR}) abstraction levels. The conversion process for this convolution kernel from \texttt{linalg} to \texttt{cinm} involves rewriting convolutions with an \texttt{im2col} (lines 1 to 7) operation followed by \texttt{matmul} (lines 8 and 9) and \texttt{expand} (lines 10 and 11) operations. The \texttt{linalg.matmul} operation is replaced with a \texttt{cinm.gemm} operation (lines 8 and 9).}
        The \cinm{} IR is subsequently lowered to
        \cim{}, \cnm{}, \texttt{affine} or a combination thereof.

\begin{table*}
\scriptsize
\caption{The \texttt{cinm} dialect operations. \(T\) marks any shaped type, \(S\) a scalar type, and \(E\) an appropriate enumeration type.}
\begin{center}
\begin{tabular}{>{\ttfamily}p{0.30\textwidth}p{0.20\textwidth}p{0.35\textwidth}cc}
\toprule
\sffamily Operation & Type & Description & CIM & CNM \\ \midrule

\verb|cinm.{add,sub,mul,div,min,max}(%lhs, %rhs)| & \(T \times T \to T\) & Element-wise arithmetic & \cmark & \cmark \\
\verb|cinm.{and,or,xor,not}(%lhs, %rhs)| & \(T \times T \to T\) & Element-wise bit-wise logic & \cmark & \cmark \\
\verb|cinm.gemv(%lhs, %rhs)| & \(S^{m\times n} \times S^n \to S^m\) & Matrix-vector product & \cmark & \cmark \\
\verb|cinm.gemm(%lhs, %rhs)| & \(S^{m\times k} \times S^{k\times n} \to S^{m\times n}\) & Matrix-matrix product & \cmark & \cmark \\
\verb|cinm.transpose(%in, %perms)| & \(S^n \times \mathbb{N}^n \to S'\) & Transposition & \xmark & \cmark \\
\verb|cinm.{histogram,majority}(%in)| & \(S^{n} \to S^{k}\) & Histogram and bit-wise majority & \xmark & \cmark \\
\verb|cinm.topk(%in, %k)| & \(S^{n} \times \mathbb{N} \to S^{k} \times \mathbb{N}^k\) & Finds \(k\) largest values \& their indices & \xmark & \cmark \\
\verb|cinm.simSearch #E, #k (%in1, %in2)| & \(E \times \mathbb{N}^k \times S^{n} \times S^{n} \times \mathbb{N} \to S^{k} \) & Finds \(k\) most similar values \& their indices with metric \(E\) & \cmark & \cmark \\
\verb|cinm.mergePartial #op #dir (%lhs, %rhs)| & \(E \times D \times T \times T \to T\) & Hardware-defined operation that merges partial results of \(E\)& \cmark & \cmark \\
\verb|cinm.popCount(%in)| & \(T \to \mathbb{N}\) & Counts \(1\)s in a bit vector & \cmark & \xmark \\
\verb|cinm.reduce #op (%in)| & \(E \times S^{n} \to S\) & Performs reduction in group \((S, E)\) & \xmark & \cmark \\
\verb|cinm.scan #op (%in)| & \(E\times S^{n} \to S^{n}\) & Performs inclusive scan in group \((S, E)\) & \xmark & \cmark \\

\bottomrule
\end{tabular}
\end{center}
\label{tab:cinm-ops}
\end{table*}

\begin{figure}[tbh]
                \subfloat[\texttt{linalg} IR for 2D convolution.]{%
                \begin{minipage}{0.9\columnwidth}
                \centering
\begin{lstlisting}[numberstyle=\tiny\color{codegray},numbers=left, language=mlir,basicstyle=\tiny\ttfamily,linerange={1-4},breaklines=true,postbreak=\mbox{\textcolor{red}{$\hookrightarrow$}\space},frame=tb,linewidth=\columnwidth]
%conv = linalg.conv2d_nhwc_hwcf ins(%img, %flt:
  tensor<1x128x128x3xi16>, tensor<3x3x3x8xi16>) outs
  (%bias:tensor<1x126x126x8xi16>)->tensor<1x126x126x8xi16>
\end{lstlisting}
                \end{minipage}%
                \label{fig:linalgIR}
                }


                \subfloat[\cinm{} IR for the 2D convolution kernel, rewritten as GEMM.]{%
                \begin{minipage}{0.9\columnwidth}
                \centering
\begin{lstlisting}[numberstyle=\tiny\color{codegray},numbers=left, language=mlir,basicstyle=\tiny\ttfamily,linerange={9-19},breaklines=true,postbreak=\mbox{\textcolor{red}{$\hookrightarrow$}\space},frame=tb,linewidth=\columnwidth]
%cbuf= linalg.generic #im2col_traits
 ins(%img: tensor<1x128x128x3xi16>)
 outs(%init: tensor<1x126x126x3x3x3xi16>) {
  ^bb0(%arg0: i16, %arg1: i16): linalg.yield %arg0 : i16
 } -> tensor<1x126x126x3x3x3xi16>
%im2col= tensor.collapse_shape %cbuf [[0,1,2],[3,4,5]]
    : tensor<1x126x126x3x3x3xi16> into tensor<15876x27xi16>
%rbuf= cinm.op.gemm %arg0, %flt :
      tensor<15876x27xi16>, tensor<27x8xi16>
%conv= tensor.expand_shape %rbuf [[0,1,2],[3,4,5]]
    : tensor<15876x8xi16> into tensor<1x126x126x8xi16>
\end{lstlisting}
                \end{minipage}%
                \label{fig:cinmIR}
                }
                \caption{Convolution kernel at different abstractions.}
                \label{fig:proglowering}
        \end{figure}
       

        In this work, we focus on the
        mechanism enforcing these mapping decisions and do not implement policy
        automation. We provide a sound infrastructure that eases the exploration
        process (automated or by the user), showing that this rewriting can be achieved at
        the \texttt{cinm} level. The development of cost models and search mechanisms
        is orthogonal to \cname{} and is left to future research
        when more reference points for comparison will be available.


        \subsubsection{The \cnm{} dialect}
        \label{sec:cnm}
        \rev{Every CNM architecture shown in \autoref{fig:taxonomy} has unique features that make it superior to others for certain workloads.
        In general, the performance efficiency is strictly determined by the parallel processing units (PUs) and data locality, i.e., in WRAM for UPMEM and in GRF for FIMDRAM. 
        These aspects are abstractly represented by the \cnm{} abstraction.}
        This intermediate dialect abstracts over the common features of CNM architectures, which use a tightly coupled hierarchy of memory and compute elements.
        It aims to provide a low-complexity transition from parallel workloads to memory-distributed programs that can guarantee access patterns and mappings.


         \begin{figure}[t]

                \subfloat[\texttt{cnm} IR for convolution.]{%
                \begin{minipage}{0.95\columnwidth}
                \centering
\begin{lstlisting}[numberstyle=\tiny\color{codegray},numbers=left,language=mlir,basicstyle=\tiny\ttfamily,linerange={1-28},breaklines=true,postbreak=\mbox{\textcolor{red}{$\hookrightarrow$}\space},frame=tb,linewidth=\columnwidth]
#scatter_map = affine_map<(d0, d1) ->
  (d0 floordiv 16, d1 floordiv 16, d0 mod 16, d1 mod 16)>
...
%C_pad = scf.for %o0 = %cst0_i to %cst15888_i step %cst128
 iter_args(%in_result = %in) -> tensor<15888x16xi16> {
 %A_tile = tensor.extract_slice %A_pad[%o0, %cst0][128,32][1,1]
  : tensor<15888x32xi16> to tensor<128x32xi16>
 %wg = cnm.workgroup [8 2] {cnm.physical_dims=["dpu","thread"]}
 %A_buf = cnm.alloc() for %wg { cnm.physical_space = "global" }
  : !cnm.buffer<16x16xi16, level 0> for !cnm.workgroup<8x2>
  ...
 %sc_a_tok = cnm.scatter %A_tile into %A_buf[#scatter_map]
  of %wg : tensor<128x32xi16> into
    !cnm.buffer<16x16xi16, level 0> of !cnm.workgroup<8x2>
 ...
 %e_tok = cnm.launch %wg (%A_buf, %B_buf, %C_buf:
    !cnm.buffer<16x16xi16, level 0>, ...) {
  ^bb0(%A_space: memref<16x16xi16>, %B_space: ...):
   scf.for %arg2 = 0 to %cst16 {
    %0 = memref.load %A_space[%arg0, %arg1] : memref<16x16xi16>
    %1 = memref.load %B_space[%arg1, %arg2] : memref<16x16xi16>
    ... }
  } : !cnm.workgroup<8x2>
  %C_tile, %g_tok = cnm.gather %C_buf[#scatter_map] of %wg
    : !cnm.workgroup<8x2> into tensor<128x32xi16>
  ...
  scf.yield %out_result : tensor<15888x16xi16> }
...
\end{lstlisting}
                \end{minipage}%
                \label{fig:cnm-ir}
                }

                \hfill

                \subfloat[\texttt{cim} IR for convolution.]{%
                \begin{minipage}{0.95\columnwidth}
                \centering
\begin{lstlisting}[numberstyle=\tiny\color{codegray},numbers=left,language=mlir,basicstyle=\tiny\ttfamily,linerange={1-26},breaklines=true,postbreak=\mbox{\textcolor{red}{$\hookrightarrow$}\space},frame=tb,linewidth=\columnwidth]
...
%C_pad = scf.for %o0 = %cst0_i to %cst15888_i step %cst128
 iter_args(%in_result = %in) -> tensor<15888x16xi16> {
 %bias = arith.constant dense<0.0> : tensor<16x16xi16>
 %C_tile = scf.for %o1 = %cst0_i to %cst128 step %cst16
  iter_args(%in_tile = %bias) -> tensor<16x16xi16> {
   %A_tile = tensor.extract_slice %A_pad[%o0, %o1][16,16][1,1]
     : tensor<15888x32xi16> to tensor<16x16xi16>
   %B_tile = tensor.extract_slice %B_pad[%o1, 0][16, 16][1, 1]
     : tensor<32x16xi16> to tensor<16x16xi16>
   %id = cim.acquire : cim_id
   %C_tile = cim.execute(%id: cim_id, %A_tile :
     tensor<16x16xi16>, %B_tile: tensor<16x16xi16>){
       %out = cinm.gemm %A_tile, %B_tile :
         tensor<16x16xi16>, tensor<16x16xi16>
       cim.yeild %out: tensor<16x16xi16> }
   cim.release %id : cim_id
   %out_tile = arith.addf %in_tile, %C_tile
     : tensor<16x16xi16>
   scf.yield %out_tile : tensor<16x16xi16> }
  %out_result = tensor.insert_slice
    %C_tile, %in_result[%o0, 0][16, 16][1, 1]
    : tensor<16x16xi16> into tensor<15888x16xi16>
  scf.yield %out_result : tensor<15888x16xi16> }
...
\end{lstlisting}
                \end{minipage}%
                \label{fig:cim-ir}
                }
                \caption{\cnm{} and \cim{} IRs for 2D convolution.}
                \label{fig:cnmcim-IR}
        \end{figure}

        \autoref{tab:cnm-ops} presents the set of operations supported by the \cnm{} abstraction. We separate the host and the device codes and represent device resources (memory/compute) with workgroups.
        \rev{\autoref{fig:cnm-ir} shows the \cnm{} IR of the convolution example in Figure~\ref{fig:linalgIR}, after applying tiling at the \cnm{} abstraction considering a tile size of $16\times16$, demonstrating workgroup creation, its mapping onto the CNM system resources, and the launching of its execution.}

\begin{figure}[htb]
            \centering
            \includegraphics[width=0.8\columnwidth]{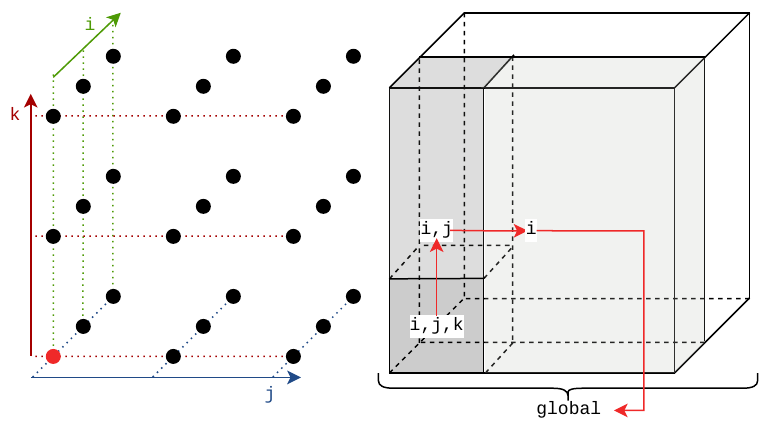}
            \caption{A \cnm{} workgroup with dims \((i,j,k)\). The red PU on the left lives in the 4 non-aliasing memory spaces shown on the right. The red path shows the tree hierarchy of accesses.}\label{fig:cnm-device-model}
        \end{figure}

        \rev{
        A workgroup is a logical address space that arranges memory resources in a tree, with PUs at its leaves, see \autoref{fig:cnm-device-model}. The dots in the left grid show logical PUs, while the different memory spaces on the right correspond to different memory hierarchy levels. For instance, the $(i,j,k)$ space could represent the local-private memory, the $(i,j)$ column, a shared memory shared by some PUs, and the $(i)$ space another memory shared by a larger set of PUs, all backed by the global memory.
        }
        ~\\
        \rev{
        PUs in a workgroup run in parallel, and lowering will eventually map slices of them to concurrent execution units of the target device.
        This is similar to CUDA, except that the workgroup dimensions can be chosen freely and that memory is intrinsically bound to levels in the tree.
        A PU can only access memory along its path to the root (the global memory), and \cnm{} forbids the user from manipulating any memory directly.
        Instead, \texttt{allocate} is used to obtain an opaque \texttt{buffer}, which is then copied to and from using \texttt{scatter} and \texttt{gather}.
        Inside a workgroup \texttt{launch} operation, the opaque buffers turn into regular \texttt{memref}s to the device memory.
        This pattern ensures that memory can be moved in the tree without invalidating the program.
        Note that physical memory is not required to have the same hierarchy, as a buffer can always be moved safely to a parent level (at a latency cost).
        For instance, in UPMEM, MRAM is technically global but is partitioned for exclusive access by the code generator.
        }

\begin{figure}[htb]
            \centering
            \includegraphics[width=\columnwidth]{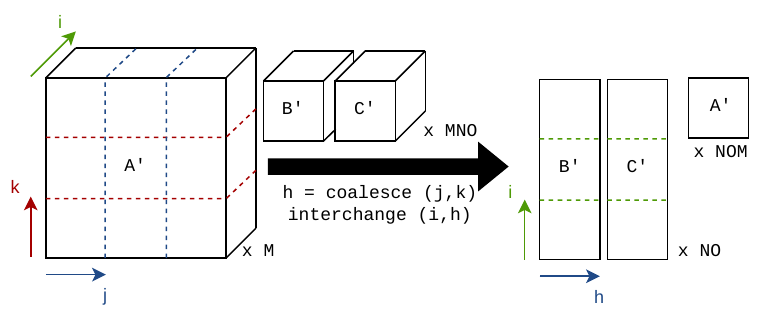}
            \caption{An example workgroup transform for \(x_{ijk} = A_{ir}B_{rjk} + C_{jk}\) from \((i,j,k)\) to \((h,i)\), showing the buffer nodes storing the working set.}\label{fig:cnm-esn-dpu}
        \end{figure}

        \rev{
        \cnm{} allows us to map parallelism inherent in an algorithm to concurrency on the device without relying on implementation details, like thread IDs (which are emulated if needed as with UPMEM) and implicit memory sharing.
        Suppose we want to implement the Einsteinian tensor expression \(x_{ijk} = A_{ir}B_{rjk} + C_{jk}\) with \texttt{A[M,P]}, \texttt{B[P,N,O]}, \texttt{C[N,O]}.
        We can immediately assume \(i,j,k\) as our parallel workgroup domain over \texttt{[M,N,O]}.
        We can also tell that each PU will have to have access to some slices \texttt{A'[P]}, \texttt{B'[P]}, and \texttt{C'[]} to avoid reduction dependencies across PUs.
        We can interchange, coalesce, and split dimensions of the workgroup freely since the PUs are independent, and this will not change their working set \texttt{A',B,' C'} either.
        When doing so, the compute remains unchanged, but the buffers commute, changing the total device memory required and the amount of scalars copied.
        \autoref{fig:cnm-esn-dpu} shows an example of such a transform, which changes the memory footprint from \(M\left(P+NO\left(P+1\right)\right)\) to \(NO\left(MP+P+1\right)\), which is advantageous for large \(M\).
        From here, the resulting 2D workgroup can be tiled as shown in \autoref{fig:load_partition} to utilize the device and obey physical memory restrictions.}



\begin{table}
\scriptsize
\begin{center}
\caption{The \texttt{cnm} dialect operations.}\label{tab:cnm-ops}
\begin{tabular}{p{0.33\linewidth}p{0.49\linewidth}}
\toprule
Operation &  Description \\ \midrule

cnm.allocate(\%arg, \%arg2 ) & 
Allocate workgroup on the specified CNM device. \\

cnm.launch(\%arg, \%arg ) & 
Launch the workgroup execution. \\

cnm.scatter(\%arg, \%arg2 ) & 
Move specified elements' indices of the input tensor to the destination tensor. \\

cnm.gather(\%arg, \%arg2 ) & 
Symmeterical to scatter, copy back. \\

cnm.wait(\%arg, \%arg2 ) & 
Wait to synchronize. \\
\bottomrule
\end{tabular}
\end{center}
\end{table}

        

        \subsubsection{The \cim{} dialect}
        \label{sec:cim}
        The \texttt{cim} abstraction serves the same purpose as \cnm{} but for CIM
        targets. Similar to the \cnm{} abstraction, it
        implements functions for acquiring/releasing the CIM accelerators, the data transfers,
        and the launching of the CIM kernel execution. Note that the CIM and CNM systems are fundamentally different, and so is the process of their resource allocation and management.  Since most CIM devices are nonvolatile, this abstraction also implements device locking to ensure consistent and
        permanent NVM states. \autoref{tab:cim-ops} lists the set of operations supported by this dialect while \autoref{fig:cim-ir} shows its IR \rev{for the tiled version of the running example with the tile size set as $16x16$}.

        The rationale for separating the CIM abstraction from the CNM abstraction is based on the architectural differences between the two paradigms they serve.
        However, within the CIM paradigm itself, despite varying operations and configurations,
        there are common properties in different setups, e.g., content-addressable memory (CAM)-based CIM, logic CIM, and crossbars.
        For instance, all these CIM types require a device setup before executing any meaningful operations,
        and the series of required operations are mostly similar.
        \rev{ The exact implementation of the device setup depends on the CIM target. For example, with crossbars, the controller orchestrates data flow and issues instructions in the correct sequence. The device setup involves configuring the controller and determining parameters such as device levels, the number of shared ADCs, and the write mode (open-loop or write-verify), among others. In the case of RTM, this entails configuring the data storage mode (sequential or bit-interleaved) and setting up the device for transverse read operations~\cite{hdc}, considering the transverse read distance.       
        }       
        The \texttt{acquire}
        function in CIM acquires a CIM device by first setting it up. The \texttt{read} and \texttt{write} operations facilitate data transfers and accessing/updating the data in the crossbar arrays. The \texttt{execute} function launches the execution in the CIM accelerator. Note that this supports both fixed single-function (e.g., crossbars) and multi-functions (e.g., DRAM-based Ambit) accelerators.
        The exact lowering is performed at the device abstraction level and varies
        depending on the specific technologies involved.

        \rev{Since CIM array sizes are
        \rev{fixed}, if the kernel size exceeds the array size, \texttt{cim} triggers
        the \textit{tiling} transformation to partition the input tensors based on the CIM array sizes in order to
        fit them into the array.}
        \rev{
        The \texttt{cim} dialect also implements CIM-specific optimizations. 
        Most CIM devices are NVM-based, and
        the write operations in almost all NVM technologies are costly in terms of both performance and device lifespan. 
        After applying the tiling pass, the order of the new loop nest $(i, j)$ can be interchanged to $(j, i)$ without encountering any dependency violations, as they are independent. In general, the \textit{loop interchange} is applied to enhance the memory access pattern~\cite{wolf1991loop}; however, \texttt{cim} uses this to minimize the number of writes.}

\begin{table}
\scriptsize
\caption{The \texttt{cim} dialect operations.}
\begin{center}
\begin{tabular}{p{0.32\linewidth}p{0.49\linewidth}}
\toprule
Operation &  Description \\ \midrule

cim.acquire() & 
Acquire a CIM device, returns ID. \\


cim.write(\%arg, \%arg2 ) &  
Write specified input tensor to the acquired CIM device. \\

cim.execute(\%arg, \%arg ) &  
Launch the execution on the acquired CIM device. \\

cim.read(\%arg ) &  
Read data from the acquired CIM device. \\


cim.barrier(\%arg, \%arg2 ) &  
Wait to synchronize or finish executing. \\

cim.release(\%arg ) & 
Release the device. \\

\bottomrule
\end{tabular}
\end{center}
\label{tab:cim-ops}
\end{table}


        
        \subsubsection{Device dialects}
        \label{sss:devaware}
        Device dialects in \cname{} expose a set of device-specific concepts, including:
        the set of supported operations, device attributes such as memory array or tile
        sizes in the CIM devices, and the memory hierarchy (buffers, private and shared
        memories). They apply conversion patterns to translate the \texttt{cinm}
        operators and provide an interface to the device libraries.

        ~\\
        \textbf{Memristors:}
        The \memrstr{} dialect implements the transformation passes from OCC~\cite{occ}
        for memristive devices. We extend the OCC flow to adapt the \cinm{} operations and transformations. \rev{The compulsory \emph{tiling} transformation is applied
        at the \cim{} abstraction to ensure large input buffers are divided into blocks
        that can be mapped to the
        crossbar tiles. The implementation details of the tiling transformation are given in Section~\ref{subsec:tiling}}. The tile size in the transformation, and hence the number of
        tiles are determined by the crossbar tile size. This dialect materializes the \cinm{} operations using the memristors' specific 
        primitives such as \texttt{copyTile}, \texttt{storeTile} to
        support data communication between the host and the device and operations such as \texttt{read} and
        \texttt{write} that allow performing computations and programming the
        crossbar, respectively.


              \begin{figure}[tb]
        \centering
        \subfloat[DPU workload]{%
        \scalebox{1}{
        \centering
        \includegraphics[scale=1.6]{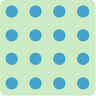}
        }%
        \label{fig:gemm_space}
        }
        \subfloat[2D (box) tiling]{%
        \scalebox{1}{
        \centering
        \includegraphics[scale=1.6]{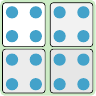}
        }%
        \label{fig:tasklets_2d}
        }
        \subfloat[Rectangular tiling]{%
        \scalebox{1}{
        \centering
        \includegraphics[scale=1.6]{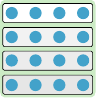}
        }%
        \label{fig:tasklets_1d}
        }
        \vspace{-.1cm}
        \caption{Various tiling shapes \cname{} implements.} 
        \label{fig:load_partition}
        \vspace{-.5cm}
        \end{figure}
        

        To enable parallel execution across multiple CIM tiles, \rev{\memrstr{} applies the
        \emph{loop unrolling} transformation on the innermost loop of the matmul kernel.
        The transformation pass takes an unroll factor, and modifies the body and loop variable of the innermost loop accordingly.
        } The partial results of individual tiles are accumulated using
        \texttt{cinm.mergePartial} as soon as they are ready.
        Finally, the \memrstr{} dialect maps all operations to the device function calls.
        All \memrstr{} operators have a one-to-one mapping with the device function calls
        exposed by the memristor devices' API. All other operations are lowered to the
        host instructions.
        
        ~\\
        \textbf{UPMEM:}
        The \upmem{} dialect features UPMEM device-specific transformation and
        optimization passes. In the architecture, each DRAM bank has an
        integrated DPU, complemented by a \SI{4}{\kilo\byte} instructions memory
        (IRAM), a \SI{64}{\kilo\byte} WRAM (scratchpad) and a \SI{64}{\mega\byte} main
        memory (MRAM). The DPUs communicate with other DPUs via the host. An
        UPMEM DIMM module integrates $M$ chips, each with $N$ DPUs.

        The \upmem{} abstraction also enables configuring the number of tasklets per
        DPU and allocating buffers in both the private WRAM and the MRAM.
        \rev{ The number of tasklets can be configured by the user. By default, \cname{} uses values that are empirically extracted for different operations and different tensor sizes. For instance, for the matmul operation, the best-performing results for large-size tensors were achieved by setting the tasklets to 16. 
        }
        For synchronization, \upmem{} introduces operations that can be ultimately mapped
        to the UPMEM \texttt{barrier\_wait} function calls for all threads.
      
      ~\\
      \textbf{Adding new devices}
        Adding a new hardware target to \cname{} requires defining a new dialect (capturing device intrinsics) and implementing the necessary conversions.
        If it supports operations that are not in the \cinm{} registered operations, which is not very common in these kinds of architectures,
        the \cinm{} dialect would need to be updated to achieve automation. However, in general, the abstractions in \cinm{}, \cim{} and \cnm{} are reused as-is for other architectures.

        \rev{For example, to support Samsung's FIMDRAM, a new device dialect needs to be added to \cname{} that contains device-specific operations, including arithmetic operations such as ADD, MAD, MUL, and MAC computing operands from different memory sources, e.g., (register file(s), bank), control operations such as JUMP, EXIT, and barrier operation. A new conversion pass needs to implemented from the \cnm{} abstraction to the new device abstraction. Since all of the operations for this target are already supported by \cinm{}, \textit{no further changes are needed} to the higher abstractions, i.e, \cinm{}, and \cnm{}.}

        ~\\\textbf{Low-level dialects}
        The optimized device-aware IRs are lowered to the low-level
        dialects common to various compilation paths. The \texttt{scf} dialect provides
        standard control flow primitives, i.e., \texttt{scf.for}, \texttt{scf.while}
        and \texttt{scf.if}. This is then lowered to the \texttt{llvm} dialect, which
        closely mirrors the LLVM IR, and can be translated to machine code.

\subsubsection{Tiling and partitioning passes}
\label{subsec:tiling}
CINM implements a generic tiling transformation using an MLIR interface that can be
triggered by the lower-level dialects. 
Operations in device-aware dialects select tile sizes based on architectural properties (e.g., crossbar size) and call this interface. 
CINM architectures perform tiling for either parallelism, improving the local memory locality, or compulsory tiling to fit operands onto CIM devices. Although the transformation is the same, it has different impacts and produces different results. 
For instance, for the matmul operation with operands of sizes $A: m\times k$ and B: $k\times n$, tiling on different dimensions produces different results and offers different tradeoffs. For instance, the box and rectangular tilings in \autoref{fig:load_partition} produce different partial results and have different localities. For CIM systems, the strategy for tiling and partitioning 
kernels must align with hardware constraints such as reliability, endurance, and other unique hardware-specific attributes. While these characteristics are inherently device-specific, they can be regarded as variables that dictate how to tile. 

        \subsection{Device cost models}
        \label{subsec:costmodel}
        For the \texttt{cinm} dialect to make an optimal device mapping decision, it
        must compare the performance of different implementations on different architectures
        considering the device constraints.
        This requires a cost model which is based on
        metrics that are comparable across devices, which is an outstanding research question.

        In our proposed flow, we designed a mechanism that can be used to leverage such
        models when available.  The \texttt{cinm} dialect declares an
        interface~\cite{interface},
        whose implementations can be registered by a device dialect at the time it is loaded.
        Considering the target hardware constraints, a \texttt{cinm} lowering conversion
        can be delegated to these interfaces to evaluate the
        cost model. When available, the appropriate selection algorithm, e.g., comparing the estimated
        ranges, will automate the mapping decision at the \texttt{cinm} level \rev{(current offloading mechanism is explained in Section~\ref{sec:cinm})}.  In
        this scenario, the \texttt{cinm} dialect will provide the advantage that the cost
        model can work on the constrained subset of interface operations defined by
        \texttt{cinm} instead of arbitrary programs.
        \rev{
            In addition to target selection, cost models can also be used to guide optimizations in \cname, similar to previous compilers for conventional architectures and accelerators~\cite{cost_locality_transform, tiramisu_cost, cost_vector, cost_spec_parallel}. They include identifying tile size and shape for a given architecture, data layout transformations~\cite{cost_locality_transform}, and other loop transformations~\cite{tiramisu_cost}. 
            }
\rev{
\subsection{Heterogeneous system setups}
\label{subsec:het-setup}
In this paper, we evaluate systems with a host CPU and a CIM/CNM device (CPU+UPMEM, CPU+memristor-crossbar) because we don't have a heterogeneous setup for evaluation. 
However, it's worth noting that heterogeneous systems utilizing conventional technologies already exist, such as Prodigy processors (combining CPU, GPU, and TPU) and Nvidia's A100 tensor core integrating multiple SMPs with GPUs for various data types and precisions. These systems are anticipated to become more prevalent with the emergence of CIM/CNM architectures. For instance, nothing prevents integrating a GPU into a UPMEM machine (CPU, GPU, UPMEM). Furthermore, advancements in System-on-Chips (SoCs) and Chiplet structures make it increasingly plausible to combine CPUs, GPUs, DPUs, TPUs, etc., once the necessary interfaces and interconnect support are in place.
~\\
The modular structure of CINM enables it to support such heterogeneous setups by allowing adding new device dialects, as detailed in Section~\ref{subsec:costmodel}, and making target selection and offloading decisions at finer-granularities, as explained in Section~\ref{sss:cinm}.
}
 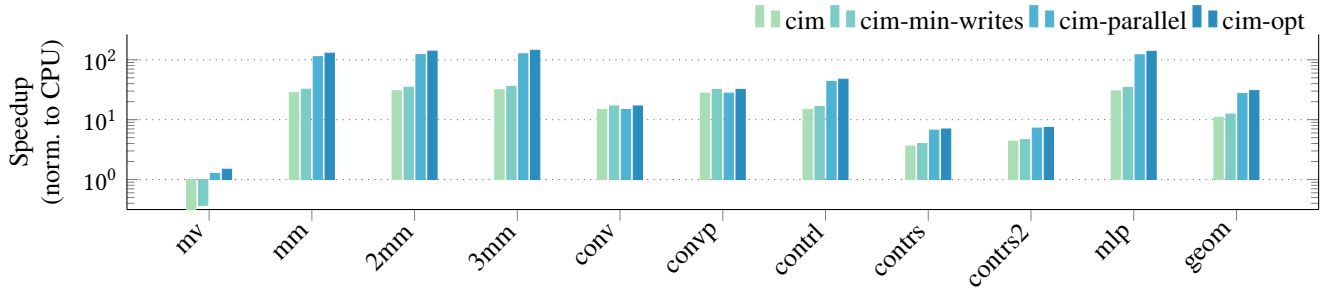
\begin{figure*}[tbh]
    \pgfplotsset{compat = newest}
    \pgfplotsset{major grid style={dotted,aluminium2!50!black}}
    \begin{tikzpicture}
    \begin{axis}
    [
            width=0.98\textwidth,
        ybar=1pt, 
        enlargelimits=0.08,
        enlarge y limits={upper, value=0.1},
        ylabel style={align=center},
        y label style={at={(-0.045,0.5)}},
        ylabel=Speedup\\ (norm. to CPU),
        legend style={draw=none, fill=none},
        bar width=3.5pt,
        legend columns=4,
        ymode = log,
        log basis y={10},
        height=0.22\textwidth,
        ymajorgrids=true,
        grid style=dashed,
        axis x line*=bottom,
        x tick label style={xshift=.0em, yshift=-.0em, rotate=45,anchor=east},
        yminorticks=true,
        legend style={at={(1,1.2)},anchor=north east},
        xlabel={},
        ytick scale label code/.code={\pgfmathparse{int(-#1)}$y \cdot 10^{\pgfmathresult}$},
        every y tick scale label/.style={at={(yticklabel cs:0.5)}, anchor = south, rotate = 90},
        symbolic x coords={mv, mm, 2mm, 3mm, conv, convp, contrl, contrs, contrs2, mlp, geom},
        xtick=data,
    ]

    \addplot+ [blind_safe_three_scheme_seven_colors_grnblu] coordinates {
    (mv, 0.315127828)
    (mm, 28.13259364)
    (2mm, 30.44479062)
    (3mm, 31.46443837)
    (conv, 14.75378786)
    (convp, 27.71897659) 
    (contrl, 14.76268904)
    (contrs, 3.613639864)
    (contrs2, 4.357384729)
    (mlp, 30.24466596) 
    (geom, 10.93375976)
    }; 
    
    \addplot+ [blind_safe_four_scheme_seven_colors_grnblu] coordinates {
    (mv, 0.370538041)
    (mm, 32.15451644)
    (2mm, 34.79727227)
    (3mm, 35.96269203)
    (conv, 16.93231837)
    (convp, 31.93122683) 
    (contrl, 16.58898514)
    (contrs, 3.973484472)
    (contrs2, 4.624143444)
    (mlp, 34.56853718) 
    (geom, 12.38542132)
    };

    \addplot+ [blind_safe_five_scheme_seven_colors_grnblu] coordinates {
    (mv, 1.260511311)
    (mm, 112.5303746)
    (2mm, 121.7791625)
    (3mm, 125.8577535)
    (conv, 14.75378786)
    (convp, 27.71897659) 
    (contrl, 43.46195007)
    (contrs, 6.67285151)
    (contrs2, 7.221782746)
    (mlp, 120.9786639) 
    (geom, 27.24409631)
    };

    \addplot+ [blind_safe_six_scheme_seven_colors_grnblu] coordinates {
    (mv, 1.482152164)
    (mm, 128.6180658)
    (2mm, 139.1890891)
    (3mm, 143.8507681)
    (conv, 16.93231837)
    (convp, 31.93122683) 
    (contrl, 47.29409609)
    (contrs, 6.963996012)
    (contrs2, 7.398629511)
    (mlp, 138.2741487) 
    (geom, 30.49440677)
    };

    \legend{cim, cim-min-writes, cim-parallel, cim-opt}
    
    \end{axis}
    \end{tikzpicture}
    \caption{Performance comparison of different CIM configurations. All results are normalized to the ARM cpu.}
    \label{fig:occ_runtime}
    \end{figure*}
\vspace{-3pt}

\section{Evaluation}
\label{sec:eval}

This section presents our experimental setup, describes our evaluated benchmarks and
gives a detailed evaluation and analysis of our generated codes and
optimizations.

\vspace{-0.2cm}
\subsection{Experimental setup}
\label{subsec:setup}
All experiments are run on an Intel Xeon CPU E5-2630 v2 @ 2.60GHz CPU having a maximum
clock frequency of \SI{3.1}{\giga\hertz}, 2 CPU sockets, 6 cores/socket, private L1 and L2 data caches sizing
\SI{384}{\kilo\byte} and \SI{3}{\mega\byte} per core, respectively, 
and a shared L3 cache of \SI{30}{\mega\byte} (2 instances). 
The machine has \SI{128}{\giga\byte} of main
memory (DRAM) with Linux Ubuntu (22.04).

For the UPMEM backend, we use a high-end real UPMEM machine with 16 DIMMs.
Each UPMEM DDR4-2400 DIMM consists of 16 PIM-enabled chips, integrating 128
DPUs.
Each DPU runs at \SI{350}{\mega\hertz}, comprises a \SI{64}{\mega\byte}
of main RAM (MRAM) and a \SI{64}{\kilo\byte} of working RAM (WRAM).
All data transfers to and between the DPUs are directed by and routed through the host.
For CIM results, we use the same setup as in OCC~\cite{occ}. All results reported in this paper are the geometric mean of ten execution runs. The
simulation environment is based on the full-system gem5 simulator~\cite{gem5} that supports memristors' based CIM accelerators. For comparison, we use the same baseline as in OCC, i.e., an in-order host ARMv8-A core with \SI{32}{\kilo\byte} and \SI{64}{\kilo\byte} instruction and data caches, respectively, and a unified \SI{2}{\mega\byte} L2 cache (cf. ~\cite{occ} for details). 
\rev{The ARM core orchestrates data/instructions to the CIM accelerator and executes all non-matmul-like operations. The execution pipeline comprises an array of CIM tiles, where each tile contains a memristor crossbar for executing analog vector-matrix multiplication. We simulated a PCM-based four-tile accelerator, each sizing $64\times64$. 
The read/write latency and energy values for the PCM devices and associated periphery are extracted from~\cite{isaac} and~\cite{gallo_compressed}.
For high-precision values, we use bit slicing by distributing computations across multiple columns, each representing a single bit. The bits are then weighted at the column output using a shift and add block.
}

\subsubsection{Benchmarks}
\label{sss:benches}
We evaluate \cname{} on two sets of benchmarks. 
\rev{
For memristor-based CIM accelerators, we are not aware of any hand-optimized benchmark suite that we could use as a baseline. Therefore, we use OCC~\cite{occ} as our baseline and compare the \cname{} generated codes with the OCC results.   
For CNM systems, we use the PriM benchmarks suite~\cite{cnm-benches}, the hand-optimized publicly available benchmarks for CNM systems, to evaluate \cname{} on UPMEM. 
}
PrIM is a collection of memory-bound workloads from various application domains, including linear algebra, database, data analytics, graph processing, bioinformatics, and image processing. 
For the non-idiomatic PriM benchmarks, no existing front-end supports translating them into an MLIR representation. 
Therefore, we resort to manual translation to handle them.
To reduce our engineering efforts, we opt not to translate and evaluate all benchmarks. 
Instead, we've chosen one benchmark from each of the 7 out of 9 categories to showcase the capabilities of our framework and the efficiency of our optimizations across a diverse range of domains. For the remaining two categories, they contain a single benchmark each in bioinformatics and sparse matrix multiplication. The translation effort for these was substantial, and we didn't observe any notable distinctions between them and other workloads.
For all non-PriM benchmarks, we start from PyTorch and use its front-end (torch-mlir) to enter MLIR and, subsequently, CINM.
All workloads in all configurations use INT32 data type. 
~\\
\textit{mm, 2mm, 3mm:} Generalized matrix-matrix multiplication \texttt{mm}, two consecutive matmuls \texttt{2mm}, and two matmuls and multiplication of their results \texttt{3mm}.
~\\
\textit{Convolution (conv)} is a compute-bound kernel dominating the execution time of most of the ML models. 
~\\
\textit{Contraction} is a generalization of matmul to N-dimensional tensors and is used in different shapes in different application domains. 
We use the examples from OCC~\cite{occ}, which were given by the indices involved in equivalent Einstein summation notation.
These are a larger contraction \texttt{contrl} \(C_{abcd} = A_{aebf} B_{dfce}\), performing two reductions, and two small contractions, \texttt{contrs1} \(C_{ab} = A_{acd}B_{dbc}\) and \texttt{contrs2} \(C_{abc} = A_{acd}B_{db}\), performing one reduction each.
~\\
\textit{MLP} is a fully connected feed-forward neural network with three layers, each consisting of a matmul followed by addition.

In addition to these benchmarks, we use benchmarks from the PrIM benchmark suite~\cite{cnm-benches}. These include vector addition (\texttt{va}), matrix-vector multiplication (\texttt{mv}), image histogram long (\texttt{hst-l}), breadth-first search (\texttt{bfs}), select kernel from databases (\texttt{sel}) and time series analysis (\texttt{ts}). 

\subsubsection{Evaluated configurations}
\label{sss:configs}
For evaluation in this section, we compare the following configurations. 

\begin{itemize}
  \item \textit{cpu-opt}: The host CPU with specification described in Sec.~\ref{subsec:setup}. All benchmarks are compiled with the Intel oneAPI DPC++/C++ Compiler 2023.1.0 on a Ubuntu host with loop-unrolled, loop-tiling, vectorization, and parallelization enabled. 
  
  \item \textit{cinm-nd}: Kernels executed in parallel on the UPMEM DPUs (having $n$ DIMMs). Each DPU uses 16 tasklets (threads). The code is generated with \cname{} flow that tiles compute kernels before offloading.

  \item \textit{cinm-opt-nd}: Code generated by \cname{} where the kernels are tiled based on WRAM size assigned to the threads, and the tiled loops are interchanged to improve the WRAM locality in the DPUs.

  \item \textit{prim-nd}: DPU-code from the PriM~\cite{cnm-benches}.

  \item \textit{cim}: Code generated by the CINM flow for the memristive CIM target. The \textit{cim} configuration applies the mandatory tiling transformation to fit compute kernels on the CIM crossbars. The \textit{cim-min-writes} configuration interchanges the tiled loops to minimize the number of write operations on the CIM devices. \textit{cim-parallel} unrolls the inner loop dimension to run multiple tiles in parallel while \textit{cim-opt} simultaneously enables all optimizations.
\end{itemize}

For CIM results, we primarily target workloads that are 
similar or can be rewritten as \texttt{mm}. 
This is because CIM devices are particularly good for mm-like 
kernels (see ~\autoref{fig:occ_runtime}, cf. \cite{cim-review, occ}) as they can execute them in constant time. 



\subsection{CIM performance comparison}
\label{subsec:callsites}
For PCM and ReRAM based CIM accelerators, the performance of the code generated by \cname{} matches the performance results presented in OCC. For completeness, we include these performance metrics obtained through the \cname{} workflow.
On average (geomean), cim outperforms the arm CPU baseline by an order of magnitude (see \autoref{fig:occ_runtime}). Note that kernels that are not \texttt{mm} and can not be rewritten as \texttt{mm} are not shown in the figure.
The cim-min-writes configuration reduces the number of writes by ~$7\times$, leading to an average performance improvement of $12.4\times$. The cim-opt delivers the best performance, i.e., $30\times$ performance gain, by combining the loop interchange and loop unrolling transformations.  

In terms of energy consumption, the \texttt{cim-opt} reduces the energy consumption by $5\times$ (geomean), compared to the host cpu. However, for some benchmarks such as \texttt{mv, conv}, the slower operands movement and little reuse on the CIM array increases the energy consumption by 30\% and 40\%, respectively compared to the baseline cpu.
\subsubsection{\cname{} evaluation on UPMEM}
\label{subsec:upmem-eval}
UPMEM systems have consistently outperformed other platforms such as CPUs, GPUs, and FPGAs, as demonstrated in numerous research studies~\cite{upmem-resources}. In this section, we extend our comparison to include a state-of-the-art CPU with all optimizations enabled. However, our primary focus here is twofold: (1) to showcase the impact of our optimizations implemented in \cname{}, and (2) to conduct a performance evaluation of \cname{}-generated code in comparison to the best-available hand-optimized codes (PriM suite) on a state-of-the-art UPMEM machine (detailed specifications in Sec.~\ref{subsec:setup}). 
Given that UPMEM is a general-purpose system capable of accelerating both ML workloads and PriM workloads, we utilize both sets of benchmarks for our UPMEM evaluation. We employ ML workloads to showcase the effectiveness of our optimizations and the PriM suite for comparative analysis.

\begin{figure}[tbh]
    \footnotesize

    \pgfplotsset{compat = newest}
    \pgfplotsset{major grid style={dotted,aluminium2!50!black}}
    \begin{tikzpicture}
    \begin{axis}
    [
        width=\columnwidth,
        ybar=0.4pt, 
        enlargelimits=0.05,
        enlarge y limits={upper, value=0.1},
        ylabel style={align=center},
        y label style={at={(-0.08,0.5)}},
        ylabel=Execution time (ms)\\ (log scale),
        legend style={draw=none, fill=none},
        bar width=2.5pt,
        legend columns=3,
        ymode = log,
        log basis y={10},
        height=0.27\textwidth,
        ymajorgrids=true,
        grid style=dashed,
        axis x line*=bottom,
        x tick label style={xshift=.0em, yshift=-.0em, rotate=45,anchor=east},
        yminorticks=true,
        legend style={at={(1,1.3)},anchor=north east},
        xlabel={},
        ytick scale label code/.code={\pgfmathparse{int(-#1)}$y \cdot 10^{\pgfmathresult}$},
        every y tick scale label/.style={at={(yticklabel cs:0.5)}, anchor = south, rotate = 90},
        symbolic x coords={mm, 2mm, 3mm, conv, contrl, contrs1, contrs2, mlp, mv},
        xtick=data,
    ]

    \addplot+ [new_one_seven_colors] coordinates {
        (mm, 180.440667)
        (2mm, 360.921667)
        (3mm, 541.355333)
        (conv, 553.194)
        (contrl, 4437.7306)
        (contrs1, 3.0334)
        (contrs2, 1400.737)
        (mlp, 2412.1268)
        (mv, 493.4876)
 
    }; 

    \addplot+ [new_two_seven_colors] coordinates {
        (mm, 124.439333)
        (2mm, 249.130667)
        (3mm, 373.55)
        (conv, 495.518)
        (contrl,3960.0326 )
        (contrs1, 0.5602)
        (contrs2, 987.362)
        (mlp, 2326.492668)
        (mv, 465.1396)
    };

    \addplot+ [new_three_seven_colors] coordinates {
        (mm,90.353667 )
        (2mm, 180.678)
        (3mm, 271.087667)
        (conv, 351.369)
        (contrl, 2301.5394)
        (contrs1, 1.714)
        (contrs2, 700.6632)
        (mlp, 1170.32134)
        (mv,247.013 )
    }; 
    
    \addplot+ [new_four_seven_colors] coordinates {
        (mm, 62.203667)
        (2mm, 124.411)
        (3mm, 186.666666)
        (conv, 276.523)
        (contrl, 1979.2632)
        (contrs1, 0.5308)
        (contrs2, 493.917)
        (mlp, 1080.23334)
        (mv, 232.3794)
    };

    \addplot+ [new_five_seven_colors] coordinates {
        (mm, 45.414)
        (2mm, 90.872)
        (3mm, 136.262333)
        (conv, 175.766)
        (contrl, 1142.278)
        (contrs1, 1.4128)
        (contrs2, 350.418)
        (mlp, 995.31267)
        (mv, 123.6274)
    }; 
    
    \addplot+ [new_six_seven_colors] coordinates {
        (mm, 31.416)
        (2mm, 62.783)
        (3mm,94.128667 )
        (conv, 138.52)
        (contrl, 989.568)
        (contrs1, 0.4448)
        (contrs2, 247.1986)
        (mlp, 989.312367)
        (mv, 116.3802)
    }; 
    \legend{cinm-4d, cinm-opt-4d, cinm-8d, cinm-opt-8d, cinm-16d, cinm-opt-16d}
    \end{axis}
    \end{tikzpicture}
    \caption{Impact of \cname{} optimizations on performance.}
    \label{fig:runtime}
    \end{figure}
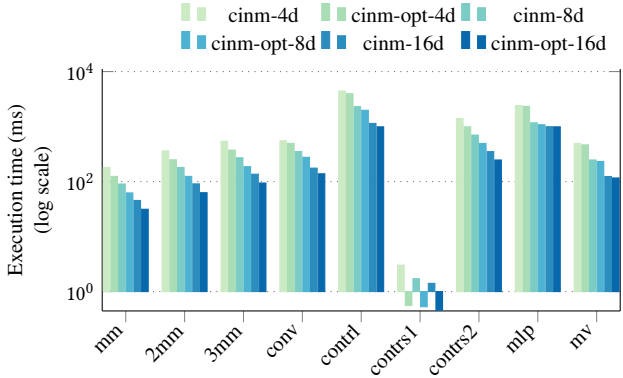
    
    
\subsubsection{Effectiveness of our device-aware optimizations}
\label{subsec:evalopt}
\autoref{fig:runtime} shows the execution time (
in \SI{}{\milli\second}) of the \cname{} generated code for the cinm and cinm-opt configurations (see Sec.~\ref{sss:configs}). 
In all benchmarks, the device-aware CINM optimizations show considerable performance gains. 
On average (geometric mean) across all benchmarks, the cinm-opt-4d, cinm-opt-8d and cinm-opt-16d configurations are $47\%$, $42\%$ and $40\%$ faster than their respective baseline cinm-nd configurations.
The speedup for \texttt{3mm} benchmark compared to \texttt{2mm} is relatively small due to the data dependencies 
of the third GEMM operation on the first two GEMM operations in \texttt{3mm}. 
The host puts the synchronization barrier after the first two multiplications 
in order to get both operands for the third multiplication before offloading it to the DPUs.

\subsection{\cname{} comparison to PriM}
\label{subsec:upmem-eval-comp}
The PriM paper~\cite{cnm-benches} extensively compares CPU and DPU systems using microbenchmarks from various domains, assessing metrics such as WRAM and MRAM bandwidth, along with DPUs' arithmetic throughput on various operations. In this section, we evaluate our generated code against their hand-optimized versions.

\begin{figure}[tbh]
\footnotesize
    \pgfplotsset{compat = newest}
    \pgfplotsset{major grid style={dotted,aluminium2!50!black}}
    \begin{tikzpicture}
    \begin{axis}
    [
        width=\columnwidth,
        ybar=0.4pt, 
        enlargelimits=0.07,
        enlarge y limits={upper, value=0.1},
        ylabel style={align=center},
        y label style={at={(-0.07,0.5)}},
        ylabel=Execution time (ms) \\(log scale),
        legend style={draw=none, fill=none},
        bar width=2.6pt,
        legend columns=4,
        ymode = log,
        log basis y={10},
        height=0.3\textwidth,
        ymajorgrids=true,
        grid style=dashed,
        axis x line*=bottom,
        x tick label style={xshift=.0em, yshift=-.0em, rotate=45,anchor=east},
        yminorticks=true,
        legend style={at={(1.05,1.35)},anchor=north east},
        xlabel={},
        ytick scale label code/.code={\pgfmathparse{int(-#1)}$y \cdot 10^{\pgfmathresult}$},
        every y tick scale label/.style={at={(yticklabel cs:0.5)}, anchor = south, rotate = 90},
        symbolic x coords={va, sel, bfs, mv, hst-l, mlp, red, ts},
        xtick=data,
    ]
    

    \addplot+ [new_one_seven_colors] coordinates {
    (va, 212.075)
    (sel, 21184)
    (bfs, 216.657996)
    (mv, 363.041)
    (hst-l, 5804.067333)
    (mlp, 1041.637)
    (red, 1063.429)
    (ts, 5242.957)
    }; 

    \addplot+ [new_two_seven_colors] coordinates {
    (va, 122.214)
    (sel, 147.374666)
    (bfs, 2423.907027)
    (mv, 460.022)
    (hst-l, 2329.814666)
    (mlp, 2521.492668)
    (red, 394.956666)
    (ts, 1566.103469)
    };

    \addplot+ [new_three_seven_colors] coordinates {
    (va, 98.878666)
    (sel, 110.458)
    (bfs, 251.318997)
    (mv, 465.1396)
    (hst-l, 623.583334)
    (mlp, 2326.492668)
    (red, 482.640666)
    (ts, 1566.103469)
    }; 
    
    \addplot+ [new_four_seven_colors] coordinates {
    (va, 61.107)
    (sel, 73.687333)
    (bfs, 4832.744941)
    (mv, 230.011)
    (hst-l, 1164.907333)
    (mlp, 1260.746334)
    (red, 197.478333)
    (ts, 757.646333)
    };

    \addplot+ [new_five_seven_colors] coordinates {
    (va, 49.439333)
    (sel, 55.229)
    (bfs, 251.318997)
    (mv, 232.3794)
    (hst-l, 311.872)
    (mlp, 1080.23334)
    (red, 120.621333)
    (ts, 811.6041875)
    }; 
    
    \addplot+ [new_six_seven_colors] coordinates {
    (va, 30.712667)
    (sel, 36.979)
    (bfs, 7801.188346)
    (mv, 115.119667)
    (hst-l, 582.392)
    (mlp, 1048.331667)
    (red, 98.860333)
    (ts, 420.052)
    };

    \addplot+ [new_seven_seven_colors] coordinates {
    (va, 25.407667)
    (sel, 27.831)
    (bfs, 251.318997)
    (mv, 116.3802)
    (hst-l, 155.929667)
    (mlp, 989.312367)
    (red, 60.578)
    (ts, 447.515031)
    };

    \legend{cpu-opt, prim-4d, cinm-opt-4d, prim-8d, cinm-opt-8d, prim-16d, cinm-opt-16d}
    \end{axis}
    \end{tikzpicture}
    \caption{Performance comparison CPU vs cinm-opt-nd and prime-nd.}
    \label{fig:runtime-comp}
    \end{figure}
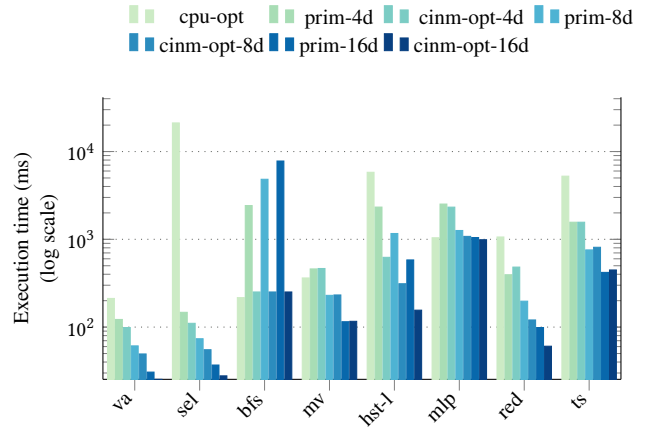

In comparison to the baseline optimized CPU configuration, both cinm-nd and prim-nd demonstrate significant performance improvements, as illustrated in Figure \ref{fig:runtime-comp}.
On average, the DPU configurations prime-4d, prime-8d, and prime-16d take $1.9\times$, $3.1\times$, and $5.1\times$, less execution time compared to the cpu-opt configuration respectively. This is primarily due to the higher number of compute units within the UPMEM systems compared to the CPU.

Comparing the prime-nd configuration to the \cname{}'s generated code (cinm-nd), the latter consistently outperforms the former except in the \texttt{mv} kernel where the performances are comparable and the \texttt{ts} kernel where prime exhibits a marginal advantage. On average, cinm-4d, cinm-8d, and cinm-16d configurations take $1.6\times$, $1.9\times$, and $2\times$, less execution time compared to the prime-4d, prime-8d, prime-16d configurations respectively.  

In most cases, the performance gains of cinm-nd over prime-nd are significant but not overwhelming. 
For instance, in the \texttt{va} kernel which has no data dependencies, the prime configuration takes \SI{122.214}{\milli\second}, \SI{61.107}{\milli\second} and \SI{30.7}{\milli\second} (absolute), respectively to compute the kernel, resulting in a speedup of up to $>7\times$ (16d) compared to the cpu-opt configuration. The cinm-nd configurations is on average $1.23\times$ better compared to the prime-nd configuration. 

However, in workloads such as \texttt{hst-l}, the \cname{}
generated codes demonstrate significantly reduced execution times compared to the prime configuration. cinm-nd takes \SI{0.623}{\second}, \SI{0.311}{\second} and \SI{0.155}{\second} for 4d, 8d and 16d configuration which are on average 3.7$\times$ less compared to their respective prime-nd configurations. The substantial gain in such benchmarks come from the better exploitation of WRAM by \cname{}. Overall, our analysis of the generated code and its comparison to the PriM implementations suggests that \cname{}'s improvements are derived from its efficient management of partial results (also dependent on the tiling size and shape, see Sec.~\ref{subsec:tiling}) and their accumulation.


Table~\ref{tab:loc} compares the applications' representation in terms of lines of code (LoC). \rev{The \textit{UPMEM (C/C++)} column shows the number of lines of code required to implement an application for both the Host and the DPUs in C/C++. The \textit{CINM (MLIR)} column shows the same application represented using the \cinm{} abstraction}. Although comparing LoCs across different programming models may be misleading, it highlights the programmability and productivity aspects of CINM. On average, idiomatic CINM is $15\times$ more concise compared to the low-level device codes. 

\begin{table}[tbh!]
\footnotesize
\caption{Comapring lines of code in CINM vs baseline.}
\centering

\begin{tabular}{rrrr}
 \toprule
  Application & CINM (MLIR) & UPMEM (C/C++) & {Reduction (times)} \\ 
 \midrule
 2mm & 19 & 184 & 10\\
 3mm & 27 & 218 & 8\\
 bfs & 29 & 315 & 11\\
 contrs2 & 14 & 200 & 14\\
 contrs1 & 14 & 197 & 14\\
 contrl & 16 & 197 & 14\\
 conv & 5 & 203 & 40\\
 hst-l & 6 & 134 & 22\\
 mlp & 58 & 109 & 4\\
 mm & 7 & 180  & 26\\
 mv & 7 & 179  & 26\\
 red & 13 & 119 & 9\\
 sel & 12 & 145 & 12\\
 ts & 25 & 172 & 7\\
 va & 7 & 101 & 14\\
 \bottomrule
\end{tabular}
\label{tab:loc}
\end{table}
 
 \section{Related work}
\rev{
Numerous software frameworks and compilation flows exist that aim to improve the programmability of CINM systems. However, most often, they target a specific architecture (memristive crossbars) or a specific domain (often machine learning).  
For instance, XLA-NDP~\cite{xla-ndp}, built on XLA~\cite{xla} with a hierarchical flow, is a compiler and runtime solution designed specifically for the NDPX~\cite{ndpx} architecture and ML training. 
PIMFlow~\cite{pimflow} targets a GPU-based CNM system (the GPU memory is PIM-enabled) for convolutional neural networks. 
It identifies the candidate CNN ONNX graph, transforms it to create inter-node parallelism across GPU and PIM, and explores different scenarios to optimize for execution time. CHOPPER~\cite{chopper} focuses on bit-serial SIMD processing using DRAM, leveraging bit-slicing techniques. Considering the CI/NM unit as a vector processor or SIMD element within the logic layer of the 3D stacked memories, PRIMO~\cite{primo} seeks patterns of vector operations in the assembly code with suitable operand sizes and substitutes them with the device APIs.
~\\
Compared to CNM systems, CIM systems have received relatively more attention. 
Ambrosi et al.~\cite{hp_memrsitor_compiler} present a software stack for a memristor-based accelerator, which converts ONNX models, enabling optimization at higher levels of abstraction.
Han et al.~\cite{polyir_memrsitor}'s compiler operates on the Polyhedral IR to identify operators suitable for offloading onto memristor accelerators.
The same objective, i.e., automatic offloading to CIM accelerators, is achieved in many other frameworks using different IRs for different CIM targets~\cite{cim_skletons, occ, tdo-cim, tc-cim, cim_mlc, rcct}. 
ComPRIMe~\cite{comprime} translates boolean and-inverter-graph-based functions into CIM-Logic instructions to offload them onto ReRAM crossbars.
}

\begin{table}
\tiny

\caption{\rev{Comparison of CI/NM compilers and software frameworks}}
\begin{center}
\begin{tabular}{>{\ttfamily}p{0.09\textwidth}p{0.005\textwidth}p{0.005\textwidth}p{0.005\textwidth}p{0.005\textwidth}
p{0.005\textwidth}p{0.005\textwidth}p{0.005\textwidth}p{0.005\textwidth}
p{0.005\textwidth}p{0.005\textwidth}p{0.005\textwidth}p{0.005\textwidth}p{0.005\textwidth}p{0.005\textwidth}p{0.005\textwidth}}
\toprule
\sffamily \textbf{Metric} &\cite{xla-ndp}& \cite{rcct} & \cite{primo} & \cite{polyir_memrsitor} & \cite{comprime} & \cite{cim_skletons} & \cite{tdo-cim} & \cite{hp_memrsitor_compiler} & \cite{tc-cim} & \cite{pimflow} & \cite{infinity_stream} & \cite{chopper} & \cite{occ,cim_mlc} & ours\\ 
\midrule
\sffamily CIM-Logic & \xmark & \cmark & \cmark & \xmark & \cmark & \cmark & \xmark & \xmark & \xmark & \xmark & \cmark & \cmark & \xmark & \cmark\\ 
\sffamily CIM-Crossbar & \xmark & \cmark & \xmark & \cmark & \xmark & \cmark & \cmark & \cmark & \cmark & \xmark & \xmark & \xmark & \cmark & \cmark\\ 
\sffamily CIM-CAM & \xmark & \xmark & \xmark & \xmark & \xmark & \cmark & \xmark & \xmark & \xmark & \xmark & \xmark & \xmark & \xmark & \cmark\\ 
\sffamily CNM & \cmark & \xmark & \xmark & \xmark & \xmark & \xmark & \xmark & \xmark & \xmark  & \cmark & \cmark & \xmark & \xmark & \cmark\\ 
\sffamily Cost model & \cmark & \cmark & \xmark & \xmark & \xmark & \xmark & \xmark & \xmark & \xmark & \cmark & \cmark & \xmark & \xmark & \cmark\\ 
\sffamily Device-agnostic input & \cmark & \cmark & \cmark & \cmark & \xmark & \cmark & \cmark & \cmark & \cmark & \cmark & \cmark & \cmark & \cmark & \cmark\\ 
\sffamily Domain-specific optimization & \cmark & \xmark & \xmark & \cmark & \xmark & \xmark & \xmark & \cmark & \xmark & \cmark & \xmark & \cmark & \cmark & \cmark\\ 
\sffamily Device-specific optimization & \cmark & \xmark & \cmark & \cmark & \cmark & \xmark & \cmark & \cmark & \xmark & \cmark & \cmark & \cmark & \cmark & \cmark \\ 
\sffamily Reusable & \xmark & \cmark & \cmark & \cmark & \xmark & \cmark & \cmark & \cmark & \cmark  & \cmark & \xmark & \cmark & \cmark & \cmark \\ 
\sffamily Hierarchical & \cmark & \xmark & \xmark & \xmark & \xmark & \xmark & \cmark & \cmark & \cmark & \cmark & \xmark & \xmark & \cmark & \cmark \\ 

\bottomrule
\end{tabular}
\end{center}
\label{tab:comp_comp}
\end{table}

\rev{Table~\ref{tab:comp_comp} compares \cname{} with existing compilers using various metrics. 
The CIM-* and CNM metrics indicate the supported device sets by the compiler. 
The Cost Model metric determines whether the compiler incorporates or allows for adding a cost model for target selection or optimization. 
The device-agnostic representation metric indicates whether the compiler requires device-specific details (such as API-calls or annotations) in the input application. 
The domain/device-specific optimization metric indicates if the compiler can apply optimizations specific to the domain or device. 
The reusable metric indicates whether the compiler can accommodate a range of devices, while the hierarchical metric denotes if it employs any form of hierarchy 
in its workflow.
~\\
Besides compilers, software frameworks exist that aim to enhance these systems' programmability. 
SimplePIM~\cite{simplepim} offers a high-level C++ library tailored for the UPMEM system. 
For CAM-based accelerators, frameworks like DT2CAM~\cite{rakka2023dt2cam} and X-TIME~\cite{xtime} have been developed to map decision trees onto TCAMs and ACAMs, respectively. The closest framework to \cname{} having a similar goal is the LLVM-IR based Infinity Stream~\cite{infinity_stream} framework that extracts the data flow graph from the C program and applies optimization and transformation passes on it for hybrid in/near memory execution. However, unlike Infinity Stream, \cname{} has multiple levels of IRs, allowing implementation of domain and device-specific optimizations and making analysis much easier compared to the LLVM IR. In terms of modularity and reusability, adding a new device in \cname{} does not require any changes to the higher level of abstractions in the CINM flow. 
}

\section{Conclusions}
\label{sec:conclusions}
We presented \cname{}, a compilation infrastructure for heterogeneous
compute-in-memory and compute-near-memory devices.  \cname{} uses MLIR rewriting
and introduces reusable abstractions and components that can be leveraged to
generalize it to other hardware targets. 
We investigated the landscape of CIM and CNM systems and presented a 
partial taxonomy of architectures along with their supported operators.
We introduced the \cinm{} abstraction that generalizes
over all CIM and CNM devices and
provides mechanisms to select a hardware target for the input kernel.  
The \cnm{} and \cim{}
dialects implement custom functions and types common to their
respective CNM and CIM devices. As concrete use cases, we presented
optimizations and code generation for memristor-based accelerators (CIM) and
the UPMEM system (CNM).  We demonstrated that by using existing and our novel
reusable abstractions, \cname{} generates code that is faster than the best-available open-source hand-optimized implementation. 

Given the emergence of CIM and CNM systems, we see \cname{} as a timely framework that empowers users to fully leverage these systems. 
It represents a significant stride towards fully automated end-to-end compilation flow that can generate highly 
optimized code for heterogeneous systems integrating 
emerging and conventional technologies. Importantly, it is already used by our collaborators. 


\section*{Acknowledgments}
This work was partially funded by the Center for Advancing Electronics Dresden (cfaed) and the German Research Council (DFG) through the HetCIM project (502388442) under the Priority Program on `Disruptive Memory Technologies' (SPP 2377), 
the AI competence center ScaDS.AI Dresden/Leipzig in Germany (01IS18026A-D).

\bibliographystyle{plain}
\bibliography{main}


\end{document}